\documentclass[11pt,ebook,onecolumn, oneside, openany]{memoir} 
\usepackage{sfimemoirarxiv}
\usepackage{tikz}
\usetikzlibrary{matrix,calc}
\usepackage{amsmath}
\usepackage{bbm} 
\newtheorem{proposition}{Proposition}

\graphicspath{{./figures/}}

\title{Greetings from a Triparental Planet}
\author{SFI Postdoctoral Fellows\\
72 Hours of Science (2019)}
\date{}

\begin{document}


\begin{frontmatter} 

{\let\cleardoublepage\clearpage 
\maketitle
}
\thispagestyle{empty}
\newpage



\thispagestyle{empty} 

\begin{center}
 
    {\kurierchap{AUTHORS}} 
    
   \vspace{0.5cm}
    
    Gizem Bacaksizlar
    
    Stefani Crabtree 
    
    Joshua Garland
    
    Natalie Grefenstette
    
    Albert Kao
    
    David Kinney
    
    Artemy Kolchinsky
    
    Tyler Marghetis
    
    Michael Price
    
    Maria Riolo
    
    Hajime Shimao
    
    Ashley Teufel
    
    Tamara van der Does
    
    Vicky Chuqiao Yang

\end{center}
\newpage


\begin{center}
    {\kurierchap{ACKNOWLEDGEMENTS}} 
    
    \vspace{0.5cm}
    
   The tradition of 72 Hours of Science originated at the Santa Fe Institute in 2016, where 15 postdocs produced, from scratch, a novel scientific paper in 72 hours. 
   
   \vspace{0.5cm}
   
   We want to thank SFI for providing the resources for this 72h retreat and especially Hilary Skolnik for her continuous encouragements and support. We are also deeply grateful to Caitlin McShea and Sienna Latham for working with us on the formatting and distribution of this unusual piece. 
    
   \vspace{0.5cm}

\end{center}

\newpage


\thispagestyle{empty} 

\begin{center}

\begin{vplace}

\textit{This is a work of speculative science. Any resemblance to actual species, civilization, or planet is purely coincidental.}

\end{vplace}

\end{center} 


\chapter*{\glyphabet{\Huge{T}}\vskip 0.25em} 
\tableofcontents* 
\thispagestyle{empty} 

\cleardoublepage

\thispagestyle{empty}

\end{frontmatter}


\chapter*{\LARGE{\glyphabet{F}}} 

{\centering\kurierchap{FOREWORD \\}}
\chaptermark{Foreword} 
\vskip\midchapskip 
\addcontentsline{toc}{chapter}{Foreword} 
\addcontentsline{toc}{section}{} 
\vskip\afterchapskip 

\pagestyle{introheadings} 


\noindent{Greetings. We are individuals from a star system far from yours. We have been assigned by our society the role of surveying the diversity of life forms across our galaxy and have been systematically visiting planets that appear suitable to sustaining life. We identified your star system as a possible life-supporting system. As such, we decided to voyage close to this location to assess its viability. During our hyperdrive jump towards your star system, we nearly collided with a small vessel made by a life form unknown to us. Deconstructing this object, we discovered within it a flat golden disk, and upon following the instructions printed on the disk, decoded the information etched into it by your society. By studying the stellar diagram on the disk, and cross-referencing this with the ratio of uranium-238 atoms found in the vessel, we realized that your planet Earth is already inhabited by a diversity of life forms. This fortuitous discovery has allowed us the opportunity to compare the life forms that exist on Earth, particularly humans, with the ones that we are familiar with in our home star system.}

The golden disk contained a great deal of information about you and your planet. You appear to understand a subset of mathematical logic and the processes that govern the physical world, which is impressive. There is great diversity in the traits of your life forms, though we did not understand much of what the images depicted. Of all of the information encoded in your disk, however, one feature surprised us most: your apparent reproductive system. Among the images, we found one which depicts two different individuals and links them to the reproduction process. This and other images suggest three facts: i) there exist two mating types (male and female); ii) one of each mating type is required for reproduction; and iii) the mating types have very different properties. We have, in fact, never observed this in any of the other life forms that we have encountered. Until now, we believed that three or more individuals were required for sexual reproduction.

This highly unusual feature of your biology convinced us to study the conditions that would lead to a triparental or a biparental system, as well as analyze in depth the consequences of triparentalism. We are sending this transcription in order to share our findings with you. We use your units of measurement, your notational system for recording mathematical ideas, and your format for recording knowledge so that you may understand us. As we approached your star system, we observed that you use electromagnetic radiation to communicate. We reverse-engineered this communication modality to access your store of knowledge and to relay this message to you. In this document, we reference knowledge from your own culture that will help you to understand the knowledge that we present to you. 

Unfortunately, our trajectory only permits a detour of 72 hours, so we are only able to give brief descriptions of our most important findings. We wanted to deposit our message at a location that will ensure that you detect it. We found an optimal signal-receiving structure, located near the village of Santa Fe, New Mexico, USA and transmitted our message to the fourteen life forms found therein.




\pagestyle{myheadings} 

\chapter[The Evolution and Consequences of Triparental Systems][]{The Evolution and Consequences of Triparental Systems}
\chaptermark{The Evolution and Consequences of Triparental Systems} 
\label{sec:evolutionandconsequences}

\addcontentsline{toc}{section}{} 
\vskip\afterchapskip 



\noindent{Based on the information contained in the Voyager, and upon examination of your scientific literature, we have determined that sexual-reproduction on Earth is biparental. In this transmission, we would like to explain the emergence and consequences of triparentalism on our home world. First, we explain the evolution of our reproductive system---that is, the conditions under which triparentalism and three self-avoiding mating types emerged as advantageous strategies for sexual reproduction. Second, we describe the biological consequences of triparental reproduction with three mating types, including the genetic  mechanisms of triparental reproduction, asymmetries between the three mating types, and infection dynamics arising from our different mode of sexual reproduction. Third, we discuss how central aspects of our society, such as short-lasting unions among individuals and the rise of a monoculture, might have arisen as a result of our triparental system. Finally, we discuss hyperdrive technology, the unified theory of physics, and the meaning of life. {\glyphabet{E}}}






\begin{mainmatter} 




\pagestyle{myheadings} 

\chapter[The Origin of Triparentalism and Mating Types][]{The Origin of Triparentalism and Mating Types}
\label{sec:origin} 

\chaptermark{The Origin of Triparentalism and Three Mating Types} 
\addcontentsline{toc}{section}{} 
\vskip\afterchapskip 



Life, both in our home star system and on Earth, has evolved sexual reproduction, such that the production of offspring requires more than one parent. However, Earth-based scientists have shown that there are potential costs of two-parent sexual reproduction, such as costs due to the so-called  ``two-fold cost of males'' \autocite{smith2017origin,smith1978evolution},  increases in disease spread, or the difficulty of coordinating multiple parents for reproduction. How then to explain the convergent evolution of a multi-parental mating strategy on both our worlds? 

This convergent evolution suggests that multi-parental mating strategies must also carry strong benefits that outweigh their costs. One theoretical model developed by an Earth-based scientist \autocite{kondrashov_selection_1982} demonstrates that sex can prevent the runaway growth of harmful genetic mutations over generations, while populations of asexual organisms undergo monotonic growth in the number of harmful mutations. This can favor the evolution of sex. 

But what is the optimal number of individuals that should combine their genetic material to produce an offspring? In Earth, outside of a small number of known exceptions \autocite{cohen2013biological,bloomfield2019triparental}, sexual reproduction occurs between two parents. In our home star system, the exclusive mode of sexual reproduction is triparental, that is, requiring three organisms. (We discuss the mechanism of how three parents typically combine genetic material in the next section.)

Species can differ not only in the number of parents required to make an offspring, but also in the number of different mating types. Mating types are subgroups of a population with different propensities to mate with each mating type. Mating types are typically self-avoiding, so that a member of a particular mating type will not mate with another member of that type. Humans have two mating types (i.e., males and females), but other organisms on Earth, such as fungi and slime molds, have many more \autocite{collins1977new}. Our own species happens to have three mating types. As in humans, our mating types are self-avoiding, meaning that each mating type does not mate with itself. Self-avoiding mating types may be beneficial because they can increase the genetic diversity of offspring. However, an organism that avoids mating with its own type also incurs increased search and coordination costs, and can generally miss out on mating with same-type but otherwise suitable mates \autocite{bull_combinatorics_1989,iwasa_evolution_1987}. 

We investigated the ecological differences between our home star system and the planet Earth that could have driven the evolution of different numbers of parents and mating types. We built on the model developed by Earth-based scientist \citet{kondrashov_selection_1982}, and extended by \citet{perry2017sex}, to examine how differences in mutation rates and coordination costs favor the emergence of systems of sexual reproduction with different numbers of parents.

\section{Model of Emergence of Triparental Systems}

To understand the origin of our triparental system, it is  helpful to first consider the benefit of multiparental (i.e., sexual) mating systems in general. One  major problem with asexual reproduction is that harmful mutations can accumulate continually over generations. The reason for this is as follows. When a mutation occurs in the genome of an asexual offspring, it is more likely that it introduces a novel harmful change rather than reversing a previously-acquired harmful change \autocite{muller1932some}. Among other advantages \autocite{kondrashov1993classification}, sexual reproduction can dramatically increase the variability found among offspring, including variability in the number of harmful mutations: some offspring will have many more harmful mutations than their parents, and some will have many fewer. Natural selection can then exploit this increased variation to select for offspring with fewer harmful mutations.

Based on these observations, Earth-based scientists created a simple model that explains the benefit of biparental mating for removing harmful mutations~\autocite{kondrashov_selection_1982,kondrashov_deleterious_1988}. Recently, however, Earth-based scientists realized that triparental reproduction can be even more effective than biparental reproduction at removing harmful mutations~\autocite{perry2017sex}. As we will see, the ability of a multiparental mating system to remove harmful mutations explains why triparentalism became dominant on our world.

However, these previous models of the evolution of bi- and triparental mating systems do not explain the origin of self-avoiding mating types, which are present in the reproductive systems found both on our world and on planet Earth. 

Note, however, that the benefit of sex occurs only when the parents are not genetic clones, since otherwise sexual reproduction becomes genetically equivalent to asexual reproduction. Self-avoiding mating types are one mechanism for guaranteeing that parents are not genetic clones, thus allowing for the mutation-reducing benefits of multiparental mating systems.  
The previously mentioned models~\autocite{perry2017sex,kondrashov_selection_1982}, for instance, assumed that parents are never clones of each other, which makes sense for Earth-based systems in which parents must be from different mating types (i.e., male and female). 

This assumption, however, did not hold on our planet when multiparental mating strategies first evolved. When triparentalism first emerged, it required only the presence of three parents, which were not differentiated into different  mating types--that is, any three organisms could in principle mate with each other. Early instances of multiparental reproduction involved three potential parents depositing a large number of gametes into a common pool. (Gametes are reproductive cells which can combine together to produce a new organism, such as the sperm and egg of your species.) In this pool, triplets of gametes would then come together and sexually fuse, thereby forming a new organism. In many cases, however, this scheme did not lead to the full benefits of sex, because the three fused gametes did not originate from three different parent organisms. In fact, when three organisms contribute gametes to a common pool, there is only a 2/9 probability that a randomly chosen triple of gametes will represent all three parents. In the case of a haploid organism (that is, one with only one copy of each gene), this implies a 7/9 probability that 2 or 3 of the three fused gametes would be clones of each other. In a triploid organism (which has three copies of each gene) that reproduces with crossover, gametes from the same parent will typically not be identical, but on average they will be more similar than those from a different parent. Thus, 
in the first multiparental mating systems that evolved in our home star system, sexual reproduction did not have the full mutation-reducing benefits predicted by Earth-based models sexual reproduction.

The fusion of clones can be avoided by having self-avoiding mating types. 
To develop self-avoidance, our life-forms developed a system in which organisms are differentiated into three different mating types (which, at this point, were still phenotypically undifferentiated). A mating event between three parents could happen only if the three parents belonged to three different mating types. Moreover, gamete triplets that came together would only initiate sexual fusion when all three mating types were represented in a triplet, guaranteeing that gametes from the same parent could never fuse together. This explains the origin of triparental reproduction with three self-avoiding mating types on our planet. The phenotypic asymmetry in the gametes of these three mating types emerged later, and is described in more detail in a later section.

The mating scheme described above guarantees self-avoidance, but it is not without some possible disadvantages. In particular, under this system, not all possible parent triplets can mate, but only those triplets that contain all three different mating types. As we state above, there is only a 2/9 probability that three parents chosen from a population will contain all three mating types. To illustrate this point with greater generality, imagine an $n$-parental system with self-avoiding mating types, where the set of  $n$ parents is assembled one-by-one. 
The first allowed parent can be of any mating type, the second parent can be of any $n-1$ mating types (i.e., excluding the mating type of the  first parent), the third parent can be any of the $n-2$ remaining types, and so on. Thus, the probability of selecting a feasible set of $n$ parents out of $n$ mating types, assuming a well-mixed population, is
$$p=\frac{n-1}{n}\frac{n-2}{n}\dots\frac{1}{n} = \frac{(n-1)!}{n^{n-1}}=\frac{n!}{n^n}\,,$$ 
a probability that decreases rapidly as $n$  increases (e.g., for a triparental system with three mating types, $p=2/9$). If a self-avoiding parent only has a single chance to find a set of mating partners, this cost can be very large. However, if there are many chances to find a set of compatible mating partners in a given mating period, then this cost decreases. Specifically, the probability of finding a feasible set of parents after $m$ chances is $1-(1-p)^m$.

The number of chances to mate can also be written as a kind of coordination cost $\frac{1}{m}$,  defined as the fraction of an organism’s total mating period that is required to find a feasible set of mating partners. The coordination cost ranges from 0 to 1: it is equal to 1 when only a single possible mating event can be attempted, and equal to 0 when an infinite number of mating events can be attempted in a given mating period (in this case, an acceptable set of mating partners will always be found for any finite $n$).

\begin{figure}
\centering
\includegraphics[width=0.99\textwidth]{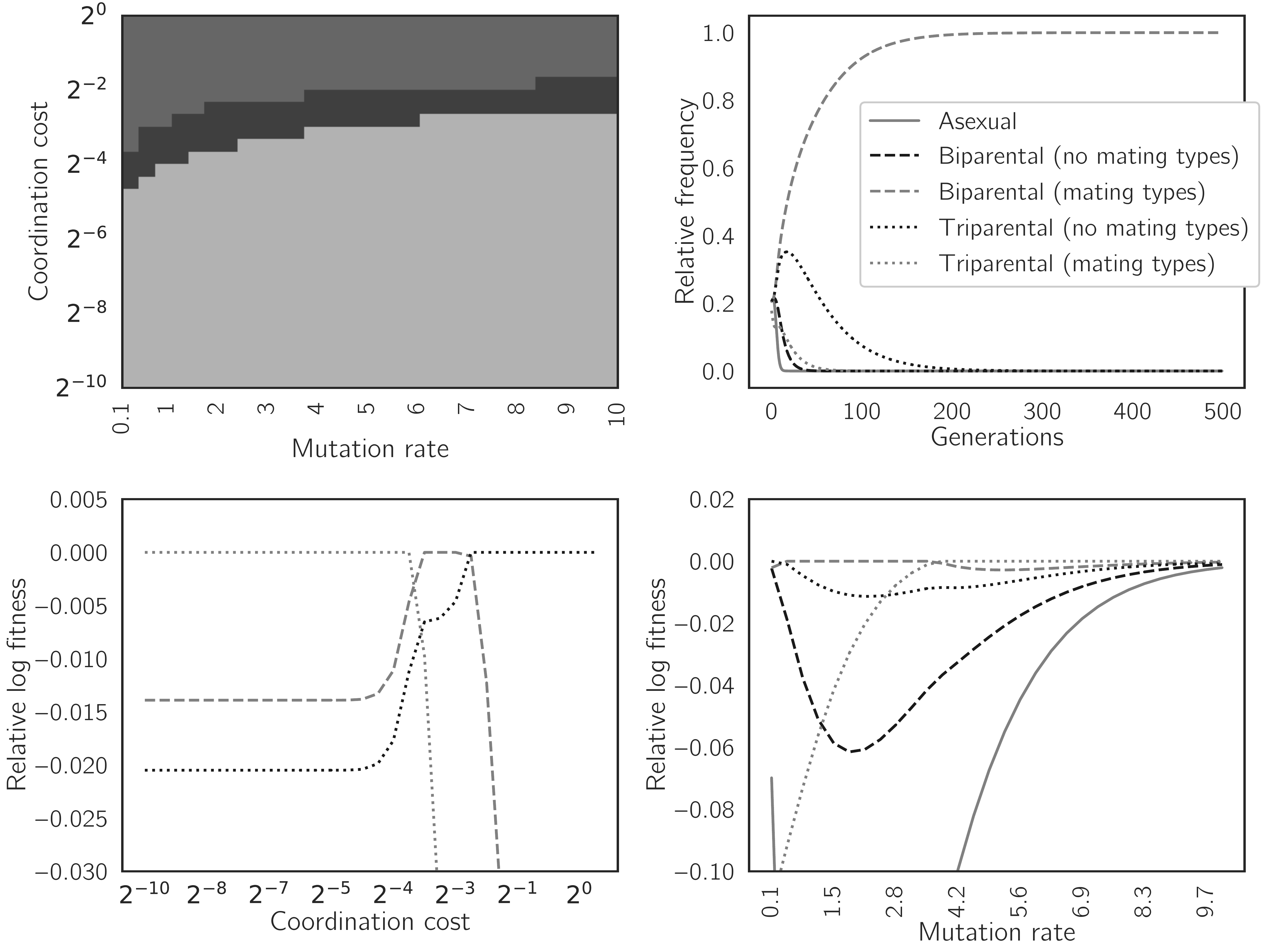}
\caption{Optimality of different mating systems {\textemdash} asexuality, biparentalism (with and without self-avoiding mating types), and triparentalism (with and without self-avoiding mating types) {\textemdash} under different conditions. 
\textbf{Top left}: Optimal mating system as a function of mutation rate (average number of mutations in an offspring) and coordination cost (fraction of mating period used to find one set of potential mates). 
Lightest gray is triparental  with mating types, darkest gray is biparental with mating types, medium gray is triparental without mating types. \textbf{Top right}: Relative frequency of different mating systems over the course of a typical evolutionary trajectory.  All mating systems start with the same frequency, and all individuals start with 0 harmful mutations.  For this run, the mutation rate was 2 and the coordination cost was $\frac{1}{8}$.  \textbf{Bottom left}: relative log fitness for different mating systems as coordination cost changes (mutation rate is fixed to $\approx 1$). \textbf{Bottom right}: relative log fitness for different mating systems as mutation rate changes (coordination cost is fixed to $\approx \frac{1}{8}$).}
\label{fig:matingsystemheatmap}
\end{figure}

To summarize, there are many different mating systems possible. The number of parents can be one (asexuality), two (biparentalism), three (triparentalism), or even larger ($n$-parentalism). There can also be different systems of mating types. Furthermore, different mating systems can carry different kinds of costs, such as the coordination cost that defines our self-avoidance scheme. 

\section{Evolution of Triparentalism and Self-Avoiding Mating Types}
Using a simple extension of the models mentioned above~\autocite{perry2017sex,kondrashov_selection_1982}, we analyzed asexuality, biparentalism, and triparentalism, both with and without self-avoiding mating types. 
We found that different kinds of mating systems are favored under different conditions, in particular different mutation rates and different coordination costs (these results are summarized in fig.~\ref{fig:matingsystemheatmap}). Further details of our model and analysis are provided in appendix~\ref{app:kondrashov} in the Supplementary Materials.

In particular, across the parameters considered by our analysis, there is a sequence of optimal mating systems going from (1) a triparental system with no mating types to (2) a biparental system with self-avoiding mating types (as in humans) to (3) a triparental system with self-avoiding mating types (the system of our home star system). Generally, a triparental system with no mating types is favored when coordination costs are high and/or when mutation rates are low, and a triparental system with self-avoiding mating types is favored when coordination costs are low and mutations rates are higher. In transitions from the former to the latter, there is always a region in which a self-avoiding biparental system becomes optimal. 

\section{Mutation and Coordination in Finite Populations}

The above model is limited to reproductive strategies with one, two, or three parents, drawn from an infinite population. This leaves a number of questions unanswered. First, does triparental reproduction dominate all $n$-parental reproductive strategies, or is it dominated under some circumstances by reproductive strategies with more parents? Second, do the benefits and costs of these mating systems differ when parents drawn from a finite, rather than infinite, population? To address these questions, we implemented Kondrashov's model of the accumulation of harmful mutations as an agent-based model in which organisms could reproduce either bi-, tri-, or quadriparentally. 

In this model, organisms belonged to independent populations of finite individuals who all reproduced either bi-, tri-, or quadriparentally. Independently of the number of parents in their mating system, species could also engage in self-avoidance, such that any one individual contributes exactly one gamete to its $n$-parental offspring. Similar to our previous model (fig.~\ref{fig:matingsystemheatmap}), self-avoidance incurs a coordination cost, which increased from 0 (no cost to self-avoidance). This cost compounded for systems with more parents, so that the overall cost of self-avoidance increased monotonically with the number of self-avoiding parents involved in reproduction.

\begin{figure}[tbh!]
    \centering
    \includegraphics[width=0.65\textwidth]{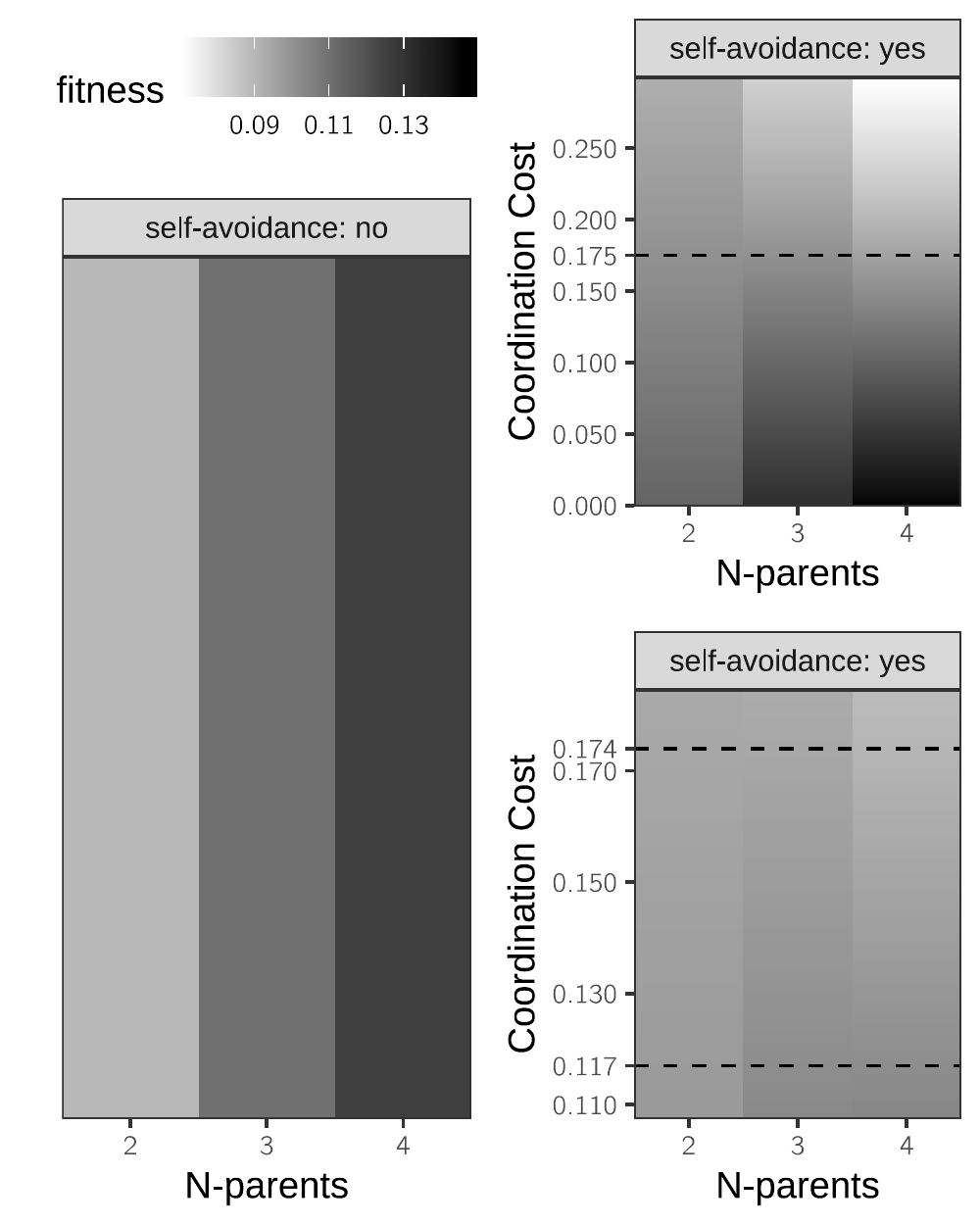}
    \caption{Relative success of $n$-parental mating systems, with and without coordination costs. \textbf{Left}: Relative success of two-, three-, and four-parent mating systems, when there is no cost to finding $n$ different parents. Color indicates relative fitness, from white (lowest fitness) to black (highest fitness). \textbf{Top right}: Relative success of each mating system when finding different parents incurs a coordination cost. Dashed horizontal line indicates the coordination cost above which biparentalism dominates all other $n$-parental mating systems. \textbf{Bottom right}: Zooming in on the range of coordination costs where triparentalism dominates other mating strategies, demarcated by the dashed horizontal lines. For coordination costs below this range, quadriparentalism dominates triparentalism.
    }
    \label{fig:tylerModel}
\end{figure}

In this model, under low coordination costs a triparental mating strategy once again dominated the biparental strategy. However, for sufficiently low coordination costs, the quadriparental strategy dominated the triparental strategy, the five-parental strategy dominated the quadriparental strategy, and so on (fig.~\ref{fig:tylerModel}, left panel). This monotonic dominance hierarchy, however, was disrupted by the cost associated with finding gametes from $n$ different individuals (fig.~\ref{fig:tylerModel}, right panels). Indeed, under reasonable assumptions about the cost of coordinating the gametes of $n$ different parents, triparentalism actually dominated both biparentalism and quadriparentalism (and thus also all forms of $n$-parentalism with $n\ne3$).

Thus, in finite populations of self-avoiding organisms in which there are costs associated with coordinating the reproduction of $n$ different parents, there is a sweet spot where the benefits of combining the genetic material of $n$ different parents outweighs the cost of coordinating the combination of those parents. Under reasonable assumptions, the optimal value of $n$ is three. It is possible that the Earth's unusual combination of low mutation rates and high coordination costs could have created an environment for biparentalism to thrive. {\glyphabet{E}}





\newpage





\pagestyle{myheadings} 

\chapter[Biological Consequences of Triparentalism and Three Mating Types][]{Biological Consequences of Triparentalism and Three Mating Types}
\chaptermark{Biological Consequences of Triparentalism} 
\addcontentsline{toc}{section}{} 
\vskip\afterchapskip 



\section{The Genetics of a Triparental Species}
\label{sec:genetics}

\begin{figure}
\centering
\includegraphics[width=0.95\columnwidth]{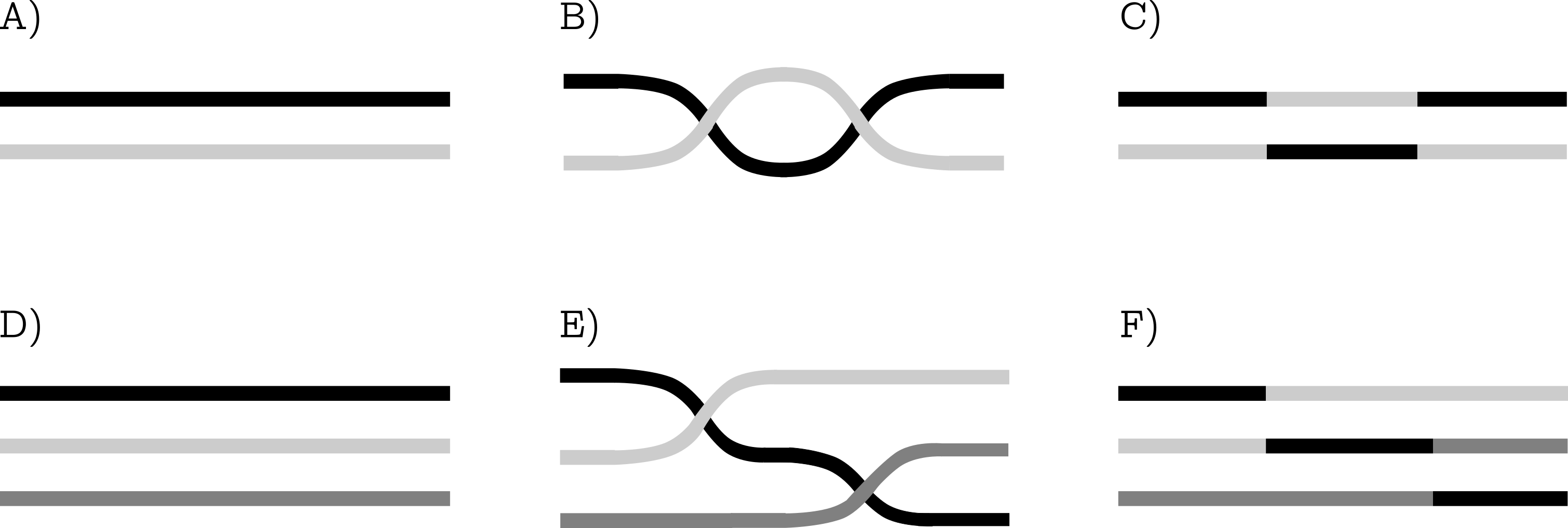}
\caption{\textbf{A)} Illustration of an Earth-based diploid genome. \textbf{B)} Chromosomal crossover during meiosis in the diploid system.  \textbf{C)} Recombined chromosomes in a diploid genome. \textbf{D)} Illustration of our triploid genome. \textbf{E)} Chromosomal crossover during meiosis in the triploid system.\textbf{F)} Recombined chromosomes in a triploid genome.}
\label{fig:crossnew}
\end{figure}

In this section, we will focus on explaining our genetics and how it compares to the genetics of your species. The details of the origin of the biochemistry of our life form, as well as the precise chemistry that forms it would take too long to include here. The crucial difference between our genetic system and yours is that our species have a triploid genome that is made from combining three haploid gametes from our three parents (each of a distinct self-avoiding mating type), as opposed to your diploid genome that is made from combining two haploid gametes from your two parents. We will refer to our genetic material as sets of double-stranded chromosomes, similar to your own genetic system, when necessary, so that you are able to understand the analogies in our descriptions. Regardless, both our genetic system and yours possess gametes: cells used during sexual reproduction to produce a new organism (e.g., sperm or egg). The details of the recombination events occurring before gamete formation are discussed in the Supplementary Materials appendix~\ref{app:genetics}, and an overview is shown in figure~\ref{fig:crossnew}.

The dynamics of our genetics and how it affects assignment of mating types is surprisingly analogous to yours but with some key differences which will be expanded upon in the Supplementary Materials appendix~\ref{app:genetics}. As discussed previously, our species is composed of three phenotypically asymmetric mating types. These mating types are distinguished by three sex-determining genotypes: `XXY,' `XYY,' and `XYZ.'

Each of these mating types produces haploid gametes (i.e., reproductive cell containing one copy of chromosomes), and one gamete of each type need to come together in order to produce a triploid offspring. Each parent thus contributes a third of its genetic material to the offspring and conversely the offpring's genome is composed of a third of each of its parent's genome.

\begin{figure}
    \centering
    \includegraphics[width = 0.9\textwidth]{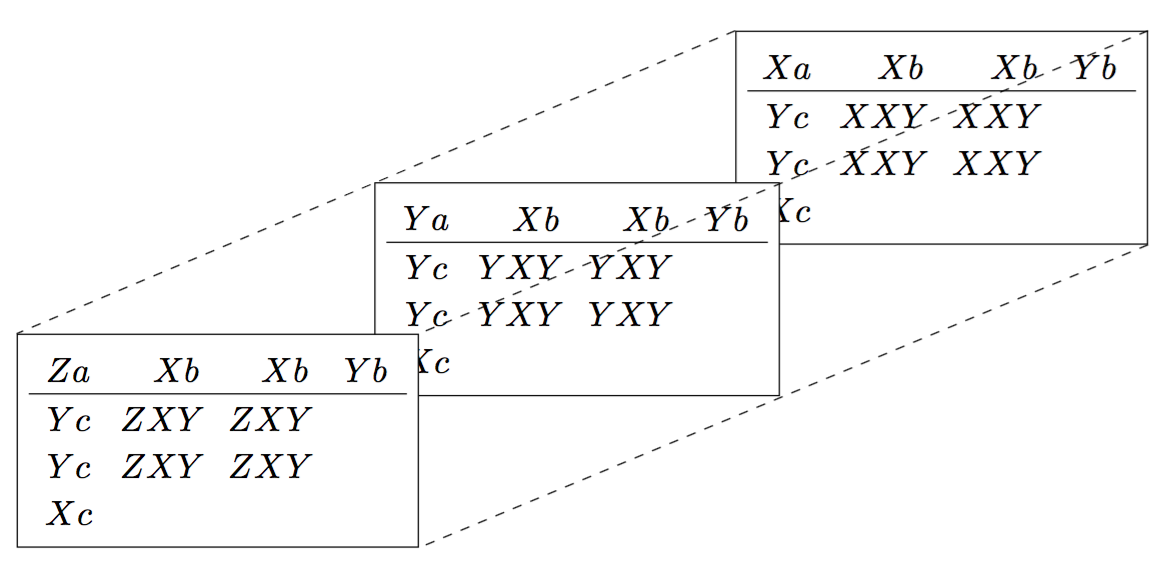}
    \caption{Partial Punnett cube of mating type assignment in our species. Letters $a$, $b$ and $c$ are used to help follow the three-mating types.}
    \label{fig:Punnett_Cube}
\end{figure}

The XYZ mating type produces the largest gamete (analogous to an egg), which can be one of three genetic types: X, Y, or Z. During reproduction, this gamete is then fertilized by the other two smaller gametes (analogous to sperm) produced by the remaining mating types. In your species, you would thus define XYZ as female and the other two as different types of males, although it is important to note that this analogy does not quite match with our sexes. The main difference to note is that the two `male' mating-types each produce only one type of gamete. The XXY genotype only produces X gametes while the XYY genotype only produces Y gametes. It follows that the female, producing all three X, Y and Z gametes, is the one that biologically determines the mating type of the offspring. We have included an example of the ``Punnett square'' that describes sex determination in the case of two mating types (table \ref{tab:punnet_human}) and its extension to illustrate this concept in our species (fig.~\ref{fig:Punnett_Cube}), illustrating this concept, and showing that all three mating-types are distributed evenly at birth.

\begin{table}
\centering
\begin{tabular}{ |c| c c| } 
\hline
& $X$ & $Y$ \\
\hline
$X$ & $XX$ & $XY$ \\ 
$X$ & $XX$ & $XY$ \\ 
 \hline
 \end{tabular}
\caption{Punnett square of mating type assignment in humans.} 
\label{tab:punnet_human}
\end{table}

\section{The Emergence of Gamete Asymmetry in Triparental Systems}

The natural history of Earth and our home planet differ in many, but not all, respects. Notably, on both Earth and our planet, mating types preceded gamete asymmetry. The preceding sections describe the evolution of triparental inheritance, mating types, self-avoidance, and genetics in the high radiation environment of our planet. In this section, we discuss the evolution of gamete size asymmetry. Additionally, we present a case study to illustrate these concepts in the context of the natural history and evolutionary ecology of the species on our planet.

On our planet, as on Earth, there is great diversity in mating systems corresponding with great diversity in fertilization strategies. Nevertheless, the modal mating system has three self-avoiding sexes. The most common pattern for gamete sizes on our planet, as mentioned in the previous section, is for two sexes to produce gametes near the minimum viable size, while one sex produces much larger gametes. However, an important minority of species have one sex that produces gametes near the minimum viable size, and two sexes that produce larger gametes. Since on Earth, individuals with small gametes are usually called males and those with large gametes females, it is perhaps tempting to associate all small gamete sexes with males and large ones with females. This is probably useful, but certainly masks a great deal of nuance, especially since our social scientists also make a similar distinction as yours between sex and gender.

To make these ideas concrete, we present both a dynamic and static model of the evolution of gamete asymmetry, both of which take three self-avoiding mating types with distinct gametes as given. 

\begin{figure}[htb!]
    \centering
    \includegraphics[width = 0.9\textwidth]{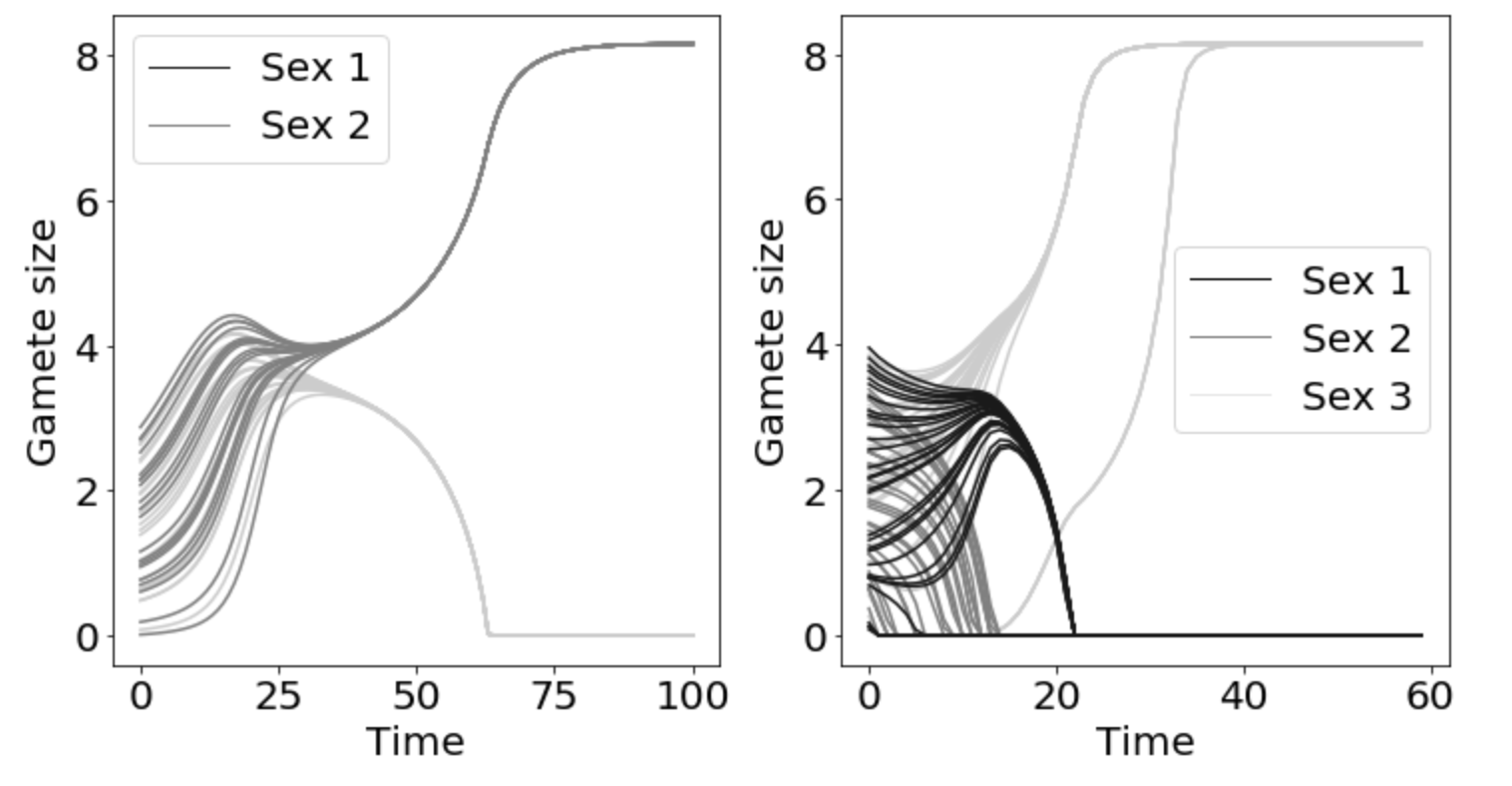}
    \caption{Simulation for the gamete size evolution. The shades of grey distinguish gametes of different sexes. Each curve indicates the evolution trajectory of one initial condition in gamete size. \textbf{Left:} in the case of two sexes, one sex evolves to have the minimal gamete size ($0$), and the other sex evolves to have a large gamete size. This recovers what happens on Earth. \textbf{Right:} in the case of three sexes, one sex evolves to have a large gamete size, and two evolves to have the minimal gamete size.}
    \label{fig:gameteSize2Panel}
\end{figure}

The dynamical model assumes a coupled system of multiple sexes evolving together. The model assumes that the gamete sizes of all sexes evolve via natural selection to increase the number of offspring that individuals have. The number of offspring depends on two components. The first is competition with members of one's own sex, where the chance of one's gamete successfully merging with that of one's mate(s) increases with the number of gametes produced. However, there is a trade-off in the total energy used for gamete production, so that producing more gametes comes at the cost of smaller gamete size. The second component is cooperation with other sexes, where the chance of the offspring being born successfully increases with the size of the two merged gametes. We perform a simulation for this model, and the results are shown in figure~\ref{fig:gameteSize2Panel}. To convince you of the model's validity, we show that the two-sex case of the model recovers what we observe is typical for Earth---one sex evolves to have a large gamete size, while the other evolves to have the minimal gamete size. In the three-sex version of the model, we show that one sex evolves to have a large gamete size, and the other two evolve to have the minimal viable gamete size. This model is described in full detail in Supplementary Materials section \ref{app:gametes}. 

The second model reaches an almost identical conclusion, albeit with different theoretical machinery. In particular, we assess the evolutionary stability of triparental inheritance with three mating types for which each mating type has exactly one distinct gamete. For the Flying-Flower species discussed in more detail below, we provide conditions so that it is an evolutionarily stable strategy (ESS) for two sexes to adopt a minimum viable gamete size, and one to adopt a large gamete size. That is, the minimal gamete size of two sexes is found to be an optimal solution, and mutating them to a larger size would not improve their fitness. When the two sexes exhibit minimal gamete sizes, the large gamete size of the other sex is also evolutionarily stable. These conditions have to do with the probability of double matings (i.e., cheating; see Supplementary Materials). So long as these probabilities are even slightly above zero, the ESS exists. With slight modifications, this model works for most species on our planet (including us), though, as we point out above, there are species with sufficiently unique mating systems such that other ESSs exist. This model is described in full detail in the Supplementary Materials section \ref{app:ess}. We now turn our attention to a case study to help illuminate these ideas. 

As on your world, gamete size asymmetry has had far-reaching consequences on the subsequent development of species that evolved to have it. This is easiest understood by loosely associating those sexes having small gametes with Earth males, and those sexes having big gametes with Earth females, though, as pointed out above, much nuance is lost by making analogies with Earth sexes. It is common on our planet, as on yours, for ``males'' of a species to evolve flashy displays as the result of sexual selection. Similarly, males of some species maintain harems and thus have evolved substantial sexual physical dimorphism.

One phenomenon that some species on our planet exhibit, that could not emerge on your planet, is frequent cooperation in seeking mates between individuals of the ``male 1'' sex and ``male 2'' sex. This is possible because these males are not competing directly with each other for genetic heritage, but do share a common interest in reproducing with ``females.'' It is common in such species that males of one sex are relatively small but specialize in costly display whereas males of the other sex are drab (both physically and behaviorally) but grow large or otherwise evolve exceptional physical characteristics, such as great agility or stamina. Clearly, the consequences of triparental inheritance and gamete asymmetry are sufficiently far-reaching that we cannot address all of those consequences here. In lieu of that, we unpack one particular case study in greater detail: the Flying-Flower genus (this is our best attempt at a literal translation of our name for this set of species).

\subsection*{A landmark case: The Flying-Flower species}

In your world, the simultaneous discover of natural selection by Charles Darwin and Alfred Wallace was a watershed moment in the biological sciences. Similarly, a watershed moment in the biological sciences on our world was the discovery that the three sexes of a single species, the Flying-Flower species, were not, as previously thought, three different species, but rather three very different sexes of the same species. Interestingly, this discovery was made independently and nearly simultaneously by \textit{three} of our scientists. Ultimately, this discovery led to the discovery of the molecular makeup of your genetic material and the mechanisms of information transfer via natural selection.

For this reason, it is the mating system analyzed in appendix \ref{app:ess} in the Supplementary Materials, despite the species not conforming to the most common mating system on our planet. While the Flying-Flower species follow the most common gamete pattern of our planet (small/small/large), mating events do not simultaneously involve all three-sexes coming together at one place to fertilize a large gamete with two small gametes. What is perhaps most notable about Flying-Flower species is that the three sexes have very different physical structures and behaviors. The so-called Flower sex ($F$) is stationary and, like plants on Earth, photosynthesizes. The other two species are the Pollinator ($P$) and Central species ($C)$. The Pollinator transfers genetic material from $F$ to $C$, but unlike with Earth pollinators, who typically receive a food reward such as nectar, its ``reward'' is that it mixes in its genetic material with the genetic material of $F$. For much of our history, our scientists assumed that the three sexes of the Flying-Flowers species were distinct species because of their great physical and behavioral differences.

Individuals of the stationary $F$ sex compete for the attention of the pollinating $P$ sex by creating flashy, decorative bundles containing their genetic material. The pollinators, $P$, do not need to engage in costly display because the more fit individuals are more adept at gathering the best bundles. Thus, their ability to gather high quality display bundles is a direct, honest reflection of their quality. The pollinators add their own genetic material to their display bundle, then display it to individuals of the central sex, $C$, who choose which bundle to accept and use for reproduction. It is rare for either $F$ or $P$ individuals to engage in substantial parental investment (e.g., they rarely protect or provide food for offspring). As a closing note for this case study, we have observed a similar phenomenon among Flying-Flower species (and other species on our planet) to what you call the Trivers-Willard hypothesis \autocite{trivers_willard1973}: high quality members of the $C$ sex will invest greater parental effort in the sex or sexes with the higher variance fitness outcomes ($F$ and/or $P$), whereas low quality $C$ individuals will invest greater parental effort in the lower variance sex or sexes.

\section{Infection Transmission Dynamics in Triparental Reproduction}

In previous sections, we discussed under what conditions triparental systems with three mating types emerge. In particular, we explained the process of genetic transmission, the emergence of asymmetrical gametes, and provide some examples of triparental species. We now move to the implication of triparental systems for infection dynamics---specifically sexually transmitted infections (STIs).

In a biparental system, STIs have the opportunity to spread if at least two individuals partake in sexual activities. Therefore, many of your models of STI contagion make the assumption that one individual spreads the infection to one other individual at a time \autocite{Eames2002, Althouse2014}. Even though, on Earth, humans do sometimes have sex with multiple individuals at once, the typical assumption is still that STIs spread from one individual to exactly one other individual at any given time. 

Some research on Earth has explored contagion models involving more than two people, however, these models still assume that the disease spreads from one individual to one other, but simply nest these interactions within highly clustered networks. While clustering sometimes hinders the propagation of diseases, it actually accelerates it if the clustering bonds are as strong as the pairing bonds see, e.g., \citet{Hebert-Dufresne2015}.

 We investigated two models for the spread of diseases via sexual transmission through small populations of biparental and triparental systems. In both parental models, we investigate how STIs could spread over a sample of 300 individuals. These models considered the effects of a wide variety of specific sexual practices by varying several parameters, e.g., average copulating probability, average commitment, average use of disease prevention methods, and average test frequency, which are assumed to be applicable to both Earth \autocite{Doherty2005} and our planet. 

In these model runs, non-coupled individuals search a local two-dimensional space out to a radius of $r$. If another single individual or couple of individuals is found within distance $r$ then those individuals choose to couple or throuple. Similarly, couples search their local space for a single to join them---if an eligible single is found, then this single and couple form a throuple.  When individuals or couples are not throupled, and have not found anyone within their local search space, they move to a new location utilizing a random walk. This behavior is continued until they discover, or are discovered by a partner or couple. It is worth making explicit that in this triparental model, a couple must continue to search the space for a third partner and may not partake in intercourse without a third. While intercourse with couples seems theoretically possible, it is held in disdain in our culture and is therefore exceedingly rare. 

At each time step, throupled groups participate in coitus, dependent on their copulating probability. In addition to copulating probability, the agents' dynamics are also dictated by a few other parameters.  For example, agents have a preference as to their frequency of STI testing and the use of protection. Furthermore, their knowledge about their own and their partners' possible infection(s) affects the intercourse dynamics. At each time step, couples and throuples also have the choice of breaking up based on their commitment threshold. Finally, while we are aware that the infection rate, as well as average time to show symptoms, are essential for the accurate modeling of disease spread, these are planet- and species-dependent, so we kept them fixed across our simulations. 

An important result of our preliminary analysis---and something which we have been aware of for quite some time---suggests that across a wide variety of parameters, triparental intercourse results in a much higher ratio of infected individuals. Because of this, early in our history we genetically modified the major STIs found on our planet so that they provided beneficial effects to us (similar to beneficial bacteria in your gut microbiota). Therefore, triparental intercourse, in addition to being necessary for reproduction, also provides a `herd immunity' effect by which our population maintains several important benefits (e.g., protection from other infectious diseases). We found that work by Earth-based researchers, also based in Santa Fe, also recently discovered that beneficial epidemics can spread through a population \autocite{berdahl2019dynamics}. This is particularly useful to us considering our society has a high degree of mixing of mating throuples in order to maximize happiness in short-term unions (see the following section). {\glyphabet{E}}




\newpage




\pagestyle{myheadings} 

\chapter[The Implications of Triparentalism for Society][]{The Implications of Triparentalism for Society}
\chaptermark{The Implications of Triparentalism for Society} 
\addcontentsline{toc}{section}{} 
\vskip\afterchapskip 



We review a few consequences of triparental systems for the larger structure of society. While there are countless consequences that lie beyond the scope of this brief discussion, here we concentrate on two different, interrelated outcomes:\ the rates of sustained social unions, and the presence or absence of cultural diversity.

\section{Matching Model: Unions and Divorce}

Across a variety of cultures, in various star systems, sustained stable unions between two or more individuals have emerged as a valued form of social relationship \autocite{Fortunato2015}. In your species, these are known as ``marriages.'' While marriage does not require or assume reproduction, it often involves restrictions on sexual relations and investments in children. On Earth, there are usually three configurations: 1) monogamous marriages, in which a person can have only one spouse at a given time; 2) polyandry, in which a female can have multiple male spouses at a given time; and 3) polygyny, in which a male can have multiple female spouses at a given time. While polygyny was historically the default union system in your species, monogamy emerged in situations that required more control of inheritance \autocite{Fortunato2015, Fortunato2010, Archetti2013}. 

Our society does not share marriage as an institution, and three-person monogamous marriages have not become a widespread norm. We suspect this is a consequence of the larger number of sexes that must be matched for reproduction in our species. From what we observed in your culture, it seems that your marriages are very stable, or at least aim to be. In your ``movies'' and ``drama series,'' we have observed that couples frequently declare ``eternal love'' and they decide to ``live happily ever after.'' 

Our institution of social union does not work that way. The union relationships among our species are typically much more short-lived than unions among humans on Earth, and most parties to union do not anticipate that it will last for the remainder of their lifetime. Our people typically repeat the process of finding mates for reproduction, getting ``divorced'' immediately from those mates, and searching for new mates. This difference is explained by the difficulty of finding the optimal mates in our triparental system. Imagine we have a population consisting of $N$ individuals of each union type. In a biparental system, you only need to find a ``soulmate'' from $N$ members of the opposite sex (or, somewhat less commonly, of the same sex), which is ridiculously easy in our ``eyes.'' In our society, on the other ``hand,'' a given individual must find their perfect ``soul-pair''---the two other individuals that form their optimal triplet---but this union must be found out of $N^2$ combinations of possible mates. Only then would members of such an optimal union marry, to use your idiom, ``till death do the three of us part.'' Of course, this happens only exceedingly rarely.

Figure~\ref{fig:married_rate} demonstrates this point via a simple model, which is described in more detail in Supplementary Materials appendix \ref{app:matchingdivorce}. In this model, people from each union type randomly meet, form a union (a ``marriage''), evaluate their quality of life (``utility''), and sustain their union so long as it improves all the partners' current situation. If we repeat this process infinitely, it converges to a stable equilibrium. As you can see, the average ``marriage'' rate is lower for societies with more sex types, and as a result the average utility tends to be lower, too.
\begin{figure}[htb!]
    \centering
    \includegraphics[width = 0.7\textwidth]{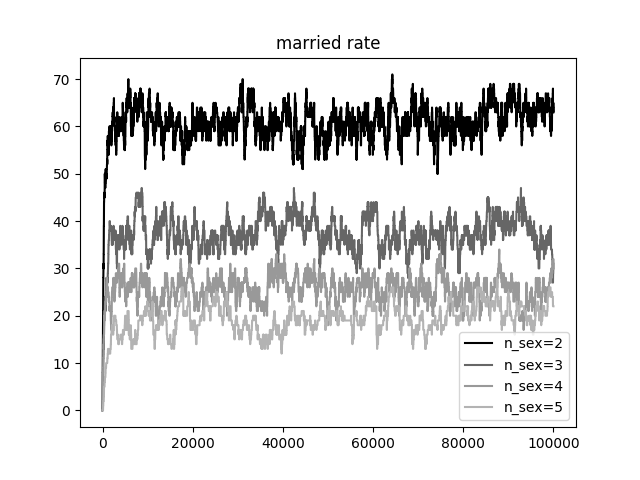}
    \includegraphics[width = 0.7\textwidth]{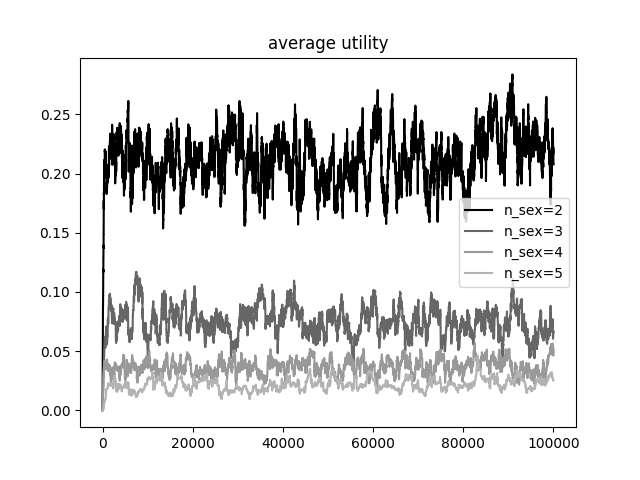}
    \caption{The ratio between the percentage of the population in a union relationship and the average utility.}
    \label{fig:married_rate}
\end{figure}

Note that this model only illustrates the relative difficulty of finding a sustainable union in systems with varying numbers of partners. In reality, we have excellent institutional and technological tools tailored to our triparental union that facilitate the creation of unions (e.g., dating platforms such as Thrindr). This makes meeting easier, but does little to change the instability of the relationships that typify our species.

Also, unions in our system are actually slightly more stable than predicted by this simple model, due both to the desire to avoid paying the search cost associated with searching for a new triplet and also to the demands of care for shared offspring. However, the fact that achieving a sustainable union has low probability has had crucial consequences in our society, particularly with regard to cultural homogeneity, as discussed in the next section.

To note, on Earth, there are examples of relationship systems that involve more than two people, and much Earth research has investigated the consequences of these systems on well-being. In societies where monogamy and polygyny occur, polygynous groups seem to lead to greater health and security for children and women \autocite{Mulder1990}. Generally speaking, on Earth, polygynous households are associated with greater wealth; whether this is due to the fact that wealthier men can have more wives, or those with more wives can aggrandize, is unclear. However, the ``polygyny-threshold model'' suggests that polygyny will develop when costs associated with sharing a husband can be offset by resource accumulation that would be difficult under monogamy \autocite{Lawson13827}. What is clear from these studies by Earth-based scientists is that there is little evidence that a three- or more-parent group would lead to a decrease in the health of either children or parents \autocite{Lawson13827, Mulder1990}. Rather, evidence from your own planet suggests that tri- or more-parental unions are associated with greater health.

\section{Cultural Mixing under Triparental Reproduction}
Data we uploaded from Earth during our fly-by strongly suggest that contemporary Earth society is still highly divided into different cultures. This strikes us as surprising; when our society reached Earth's current level of capability for inter-cultural mixing, a homogeneous culture emerged quickly. In light of the salience of biparental reproduction in our study of Earth, we chose to investigate whether our own capacity for triparental reproduction can explain, at least in part, the emergence of cultural homogeneity in our society, as compared to humans' highly diverse and stratified system of cultures.\par 

\begin{figure}[htb!]
    \centering
    \includegraphics[width=.8\textwidth]{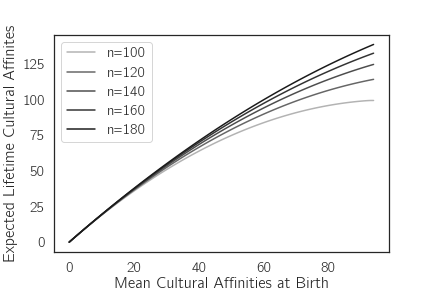}
    \caption{People born with more cultural affinities (horizontal axis) can expect to acquire an even greater number of total affinities over the course of their lifetime (vertical axis). This was true regardless of the total number of distinct cultures, $n$, in the society, indicated by the different lines. Given that individuals inherit most of their parents’ cultural affinities, this leads to a generation-to-generation increase in the mean number of cultural affinities held by members of a society.}
    \label{fig:cultureplot}
\end{figure}

Our model (provided in Supplementary Materials appendix \ref{app:culturalmixing}) shows that under triparental reproduction, individuals tend to acquire a large number of new cultural affinities over their life---and since individuals inherit most of their parents' cultural affinities, there is a rapid increase from generation to generation in the number of cultural affinities that people tend to have. At some point, the cognitive load of maintaining multiple cultural identities becomes overwhelming, even for beings like us. We hypothesize that our current cultural homogeneity emerged as a response to the explosion in cultural identities modeled above. This is due to two factors:\ the greater possibility under triparentalism for acquiring additional cultural affinities across one's life through union; and the gradual accumulation from generation to generation in the number of cultural affinities that people are born with, under the assumption that one inherits most of one's parents' cultural affinities. Indeed, as shown in figure~\ref{fig:cultureplot}, our model predicts that the number of cultural affinities that an individual has over their lifetime increases rapidly with the number of cultural affinities with which they were born.\par 

Certainly, the evidence from Earth suggests that there is a critical level of cultural complexity at which individuals become overburdened and the adoption of a single culture becomes necessary. In an ethnographic study of a highly multilingual after-school program in Los Angeles, \citet{orellana2016cultivating} find that despite speaking a wide array of languages and dialects at home, children in the program converged on a narrower range of English and Spanish dialects when speaking with each other. Data from \citet{rumbaut2013immigration} show that, while the United States is historically a country with a high rate of immigration and therefore a rich diversity of language forms, it is also a ``language graveyard,'' i.e., a country where immigrant languages increasingly fall out of use. These findings suggest that there is a level of linguistic diversity at which languages start to fall out of use, due to the need for assimilation. There is some evidence that this is due not just to the desire for assimilation into the dominant community, but also due to the cognitive load of speaking multiple languages. \citet{matras2000fusion} finds that speakers of multiple languages often develop strategies for ``fusing'' their spoken languages into one common language, so that the multilingual subject effectively speaks a single, fused language. This fusion develops as a shortcut for avoiding the cognitive load associated with maintaining fluency in two completely different languages.\par 

Language is, of course, only one aspect of culture, and the cultural homogeneity of our current society encompasses more dimensions of culture than language. Indeed, most individuals in our society speak a common language, but we also share a broader set of common cultural practices. We suspect, however, that the emergence of such cultural homogeneity began with the fusing of language, which led to a broader fusion of cultural elements communicated via common media. We also have strong evidence that this process of cultural convergence developed at a much faster rate in our society than the rate at which it is currently developing on Earth. In light of the model proposed above, we have reason to suspect that our triparental system of union and reproduction, as opposed to a two-participant system, plays a significant explanatory role with respect to the higher rate of cultural convergence that we experienced in our deep past.\par 

Indeed, our antiquarians---our best translation of a group of social scientists comprising, in your terminology, comparative linguists, archaeologists, historians, and so forth---have uncovered good, albeit indirect evidence supporting the perspective outlined above. As with your species, the adoption of agriculture was a seminal moment in our history, leading to the growth of ``urbanism'' and a rapid increase in technological innovation. Prior to this, our ancestors had spread across the continents of our world, developing distinct local cultural traditions, with physical boundaries usually demarcating the cultural boundaries between these groups. Shortly after the adoption of food production, we domesticated a beast of burden (quite unlike your horses, but serving the same functional role) that allowed more rapid movement and vastly expanded trading in material goods as well as the movement of members of our species among previously isolated locations.

This created a set of natural experiments whereby previously non-mixing groups with distinct (and locally unified) cultures came increasingly in contact. Our antiquarians have fused linguistic data with archaeological data---for the latter, we rely heavily on stylistic differences in vessels created from highly pliable soils fired at high temperatures (similar to your pottery, which we also invented)---to show that as these natural experiments played out there was a very rapid unification of cultures. This occurred far more rapidly than the most comparable examples from your history, although it was far more difficult for us to piece together our own history from artifacts on our planet.

Differences in atmospheric chemistry between our worlds means we cannot utilize radiocarbon dating, since there is not a constant production of carbon-14 from nitrogen-14 in our world. Because we cannot rely on radiocarbon dating, our scientists are much more precise at using potassium-argon decay, which for you is useful in dating fossils but for us is useful in detecting micro-minute decay-rates and provides a precise but long-term (roughly three-million-year) chronology. Calibrating this with isotopic signatures and dendrochronology from our trees roughly similar to your bristlecone pines, we are able to get a fairly good timeline, while we largely continue to operate in the regime that some of your archaeologists identify as ``cultural history.'' That being said, techniques that form the basis for (relative) dating in cultural history (e.g., seriation by artifact types) are quite reliable in our world since culture change is more rapid (aside from, it seems, the pottery changes detected by archaeologists in Pueblo culture in Colorado/New Mexico). {\glyphabet{E}}




\newpage




\pagestyle{myheadings} 

\chapter[Hyperdrive Technology, the Unified Theory of Physics, and the Meaning of Life][]{Hyperdrive Technology, the Unified Theory of Physics, and the Meaning of Life}
\chaptermark{Hyperdrive Technology, the Unified Theory of Physics, and the Meaning of Life} 
\addcontentsline{toc}{section}{} 
\vskip\afterchapskip 



We have, thus far, focused on the implications of a triparental system, from the structure of our genetic information, to the evolution of self-avoidance, emergence of gamete asymmetries, spread of sexually transmitted diseases, and social effects from marriage to cultural mixing. 

We now move on to other topics that may also be of interest to humans on Earth. We will begin with a detailed description of our hyperdrive technology that allows us to travel at the speed of light, then describe the missing pieces of your understanding of the physical world, and end with the implications of the unified theory of physics on the meaning of life.

Our hyperdrive technology is based on the well-known phenomenon of . . . 

\begin{figure}[htb!]
\centering
\includegraphics[width=1\columnwidth]{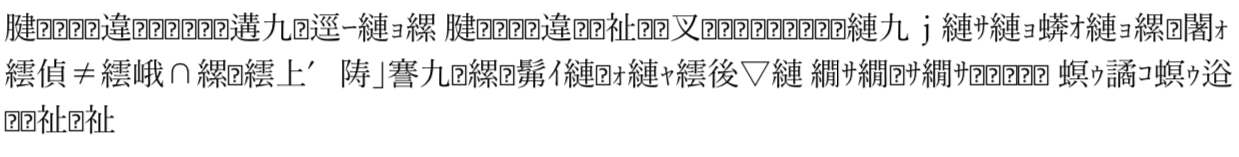}
\end{figure}


\vspace{0.5cm}

[signal lost] 

\vspace{1cm}
\newpage

\pagestyle{myheadings} 


\chapter[Afterword][]{Afterword}
\chaptermark{Afterword}

\textbf{November 4th, 2019} 

\vspace{0.5cm}

We are sorry to inform you that the transmission ended here. It appears that these extraterrestrial scientists completed their upload of their investigation of the differences between triparental and biparental mating systems and were beginning to transmit a data dump of core innovations of their culture, from their technology to the scientific knowledge and metaphysics. Unfortunately, this portion of their transmission did not arrive intact.

We have taken all of the data sent to us by our alien visitors and formatted it to the best of our ability. Extant errors within the text may be due to translation errors by the aliens, differences in logical thinking in alien cognition, or possibly transmission errors that we were not able to correct. Despite its imperfect nature, we think that it contributes valuable information, which we would like to share with the rest of the world.

We hope you enjoyed this interplanetary transmission.  {\glyphabet{E}}

\vspace{0.5cm}
\emph{—The postdocs at the Santa Fe Institute}


\includepdf{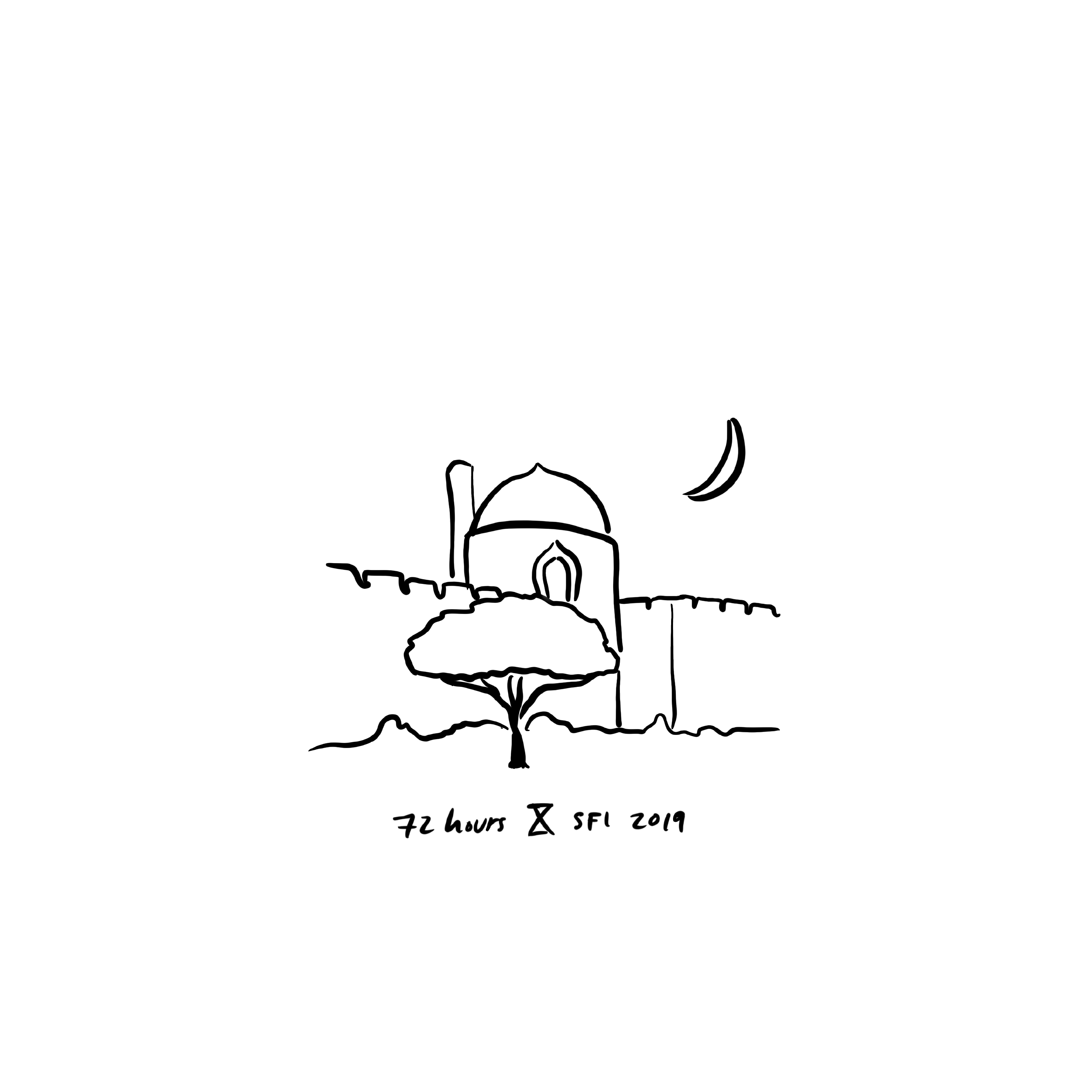}





\newpage




\pagestyle{myheadings} 

\chapter[Supplementary Materials][]{Supplementary Materials}
\chaptermark{Supplementary Materials} 
\addcontentsline{toc}{section}{} 
\vskip\afterchapskip 



\section{Model for the Origin of Triparentalism and Mating Types}
\label{app:kondrashov}

Here we describe the model described in section~\ref{sec:origin}, whose results are shown in figure~\ref{fig:matingsystemheatmap}. Our work extends a previous model, originally put forth  by the Earth-based scientist \citet{kondrashov_selection_1982}. This model was used to explore the consequence of triparental sex in a somewhat different context in \citet{perry2017sex}, which was written without knowledge of the actual existence of triparentalism in our home star system.

In this appendix, we briefly review Kondrashov's model.  Consider an infinitely large population of haploid  individuals who reproduce asexually, and let $p_{i}$ indicate the relative frequency of individuals
with $i$ harmful mutations in a given generation. Each individual has an (effectively) infinite number of loci (i.e., the genome is infinitely long,  so that there are infinitely many locations for mutations to occur), so that the probability of two mutations arising independently at the same locus is zero. 

Each organism produces gametes with  some number of new harmful mutations.  The probability that a gamete acquires $k$ new mutations is Poisson-distributed,  $\text{Pois}_\mu(k) = \frac{\mu^k e^{-\mu}}{k!}$, where $\mu$ is the average number of new mutations that will be
acquired. 
After mutations, the relative frequency of gametes with $i$
new mutations is
\begin{equation}
q_{i}=\sum_{j} p_{j}\text{Pois}_{\mu}(i-j).\label{eq:apppoisson}
\end{equation}
The gametes produce offspring that undergo
selection, where the fitness of an organism with $i$ harmful mutations
is given by $s_{i}$.  After selection, the total portion
of offspring left is $Z=\sum_{i}s_{i}q_{i}$. This means that the relative frequency of organisms with $i$ harmful mutations in the next generation is:
\begin{equation}
p_{i}'=s_{i}q_{i}/Z.\label{eq:appsel}
\end{equation}
\citet{kondrashov_selection_1982} considered
several different fitness functions, but for simplicity we use the
most basic ``threshold'' fitness function:
\[
s_{i}=\begin{cases}
1 & \text{if \ensuremath{i<K}}\\
0 & \text{otherwise}
\end{cases}
\]
Here, $K$ is a parameter that determines the maximum number of harmful
mutations that can be acquired before the organism is no longer viable.
In our simulations, we set $K=6$. 

For this simple threshold fitness function and asexual reproduction,
it can be easily shown that the stationary population frequencies
are given by $p_{i}=\delta(i,K-1)$. In this case, the population
is perched at ``the edge of the cliff'' of the fitness landscape, and the
portion of offspring preserved at the end of each selection round
is $Z=e^{-\mu}$.

We now consider what happens in the case of biparental sexual reproduction. In this scenario, parents again produce gametes with new mutations according to equation \ref{eq:apppoisson}. During mating, the gametes undergo crossover,
and the resulting offspring then undergo selection. It is assumed that parents are selected independently from an (effectively) infinite population, and that
mutations are inherited from the gametes according to a binomial sampling
process. After crossover, the probability of an offspring with $i$ harmful mutations is 
\begin{equation}
r_i=\sum_{j,k:j+k\ge i}q_{j}q_{k}\text{Binom}_{j+k,1/2}(i) \,,\label{eq:appbiparental}
\end{equation}
where $\text{Binom}_{j+k,1/2}(i)=\binom{j+k}{i}\frac{1}{2}^{i}\frac{1}{2}^{j+k-i}$ is the probability that $i$ out of $j+k$ mutations being inherited (under
a Binomial distribution with success probability $1/2$; note that under the assumption of an infinite number of loci and an infinite population, different parents
will never overlap in their mutations). Selection proceeds in the same way as described above; thus, the relative frequency of organisms with $i$ harmful mutations in the next generation is $p_{i}'=s_{i}r_{i}/Z$, where $Z=\sum_{i}s_{i}r_{i}$. 
\citet{kondrashov_selection_1982} showed that under this biparental model, the
stationary frequencies of sexual populations typically have lower
rates of harmful mutations and higher fitness.

This can be readily extended to the case of triparental reproduction
by considering a mixture of three non-clonal parents. Here, equation~\ref{eq:appbiparental}
is modified as 
\begin{equation}
r_i=\sum_{{\substack{j,k,l:\\j+k+l\ge i}}}q_{j} q_{k} q_{l}\, \text{Binom}_{j+k+l,1/3}(i) \,.\label{eq:appbiparental-1}
\end{equation}
This is similar to the model considered by \citet{perry2017sex}.

The above sexual-reproduction models assume that parents are never clones
of each other. However, when gametes are not selected uniformly at random from an infinite population of different parents, reproduction between clones can occur. For instance, consider the kind of  scenario that occurs on our world (as described in the main text): a set $n$ parents release
their gametes into a common pool, from which $n$ gametes
are then selected to make new organisms. In this case, clonal reproduction may be quite common. For biparentalism ($n=2$), half of all matings will be between clones, leading to the 
following modification of equation \ref{eq:appbiparental}:
\begin{equation}
r_i=\frac{1}{2}q_i +\frac{1}{2}\sum_{j,k:j+k\ge i}q_j q_k\, \text{Binom}_{j+k,1/2}(i)\,.
\label{eq:appbiparental-2}
\end{equation}
Reproduction between clones can be avoided using self-avoiding mating types. However, with self-avoidance,  there is the coordination cost of finding
two different parents. If only a single round of mating is allowed,
this reduces the probability of biparental mating by 1/2. More generally, if
$m$ rounds of mating are allowed, this reduces the probability of biparental
mating by $1-(1/2)^{m}$ (the probability of not failing every round), giving
\[
r_{i}=(1-(1/2)^{m})\sum_{j,k:j+k\ge i}q_{j}q_{k}\,\text{Binom}_{j+k,1/2}(i) \,.
\]
In the main text, we refer to the number $1/m$  as the ``coordination cost.''

A similar modification can be done for the triparental model. Consider
a model where three parents come together and release their gametes
into a common pool. The probability that a triple of gametes originates
from one parent is 1/9, that it originates from two parents (where
one contributes 2 gametes of the triple and the other one gamete from
the triple) is 2/3, and that it originates from three different parents
is 2/9. Thus, without self-avoidance, equation \ref{eq:appbiparental-1} is modified as
\[
r_i=\frac{1}{9}q_{i} +\frac{2}{3}\eta_{i}+\frac{2}{9}  \sum_{{\substack{j,k,l:\\j+k+l\ge i}}}
q_{j} q_{k} q_{l} \, \text{Binom}_{j+k+l,1/3}(i),\label{eq:appbiparental-1-1}
\]
where $\eta_{i}$ is the probability that an offspring has $i$ mutations, given  that one parent contributes two gametes and one parent 
contributes one gamete. This probability can be written as
\[
\eta_{i}=
\sum_{{\substack{j,k:\\j+k\ge i}}}
q_{j} q_{k} \sum_{\substack{l:0\le l \le j, \\0\le i-l\le k}}\text{Binom}_{j,1/3}(l) \,\text{Binom}_{k,2/3}(i-l).
\]
Here $q_j q_k$ is the probability that a pool is formed by a set of three parents at least one of which has $j$ mutations and at least one of which has $k$  mutations. The inner sum is the probability that the offspring has exactly $i$  mutations, given that in such a pool, a parent with $j$ mutations contributes one gamete and a parent with $k$ mutations contributes two gametes.

With can again consider the triparental system where gametes can be clones, but now with self-avoidaning mating types. Given three parents and three self-avoiding mating types, and $m$ rounds of mating, the probability that an offspring before selection has $i$ mutations is
\[
r_{i}=(1-(7/9)^{m})\sum_{{\substack{j,k,l:\\j+k+l\ge i}}}
q_{j} q_{k} q_{l} \, \text{Binom}_{j+k+l,1/3}(i) \,.
\]
(See the main text for the derivation of the factor $(1-(7/9)^{m})$.) This expression can be naturally extended to the case of an arbitrary number of parents and mating types.

As mentioned, we assumed haploid genomes.  However, the results also apply to diploid and triploid genomes, assuming mutations are dominant \autocite{kondrashov_selection_1982,kondrashov_deleterious_1988}. 
Finally, note that when comparing different mating
systems, we do not give the asexual system a twofold (or threefold)
fitness advantage. This is because such a fitness advantage only arises when the gametes
of the different mating types are asymmetric, while in this model
they are symmetric \autocite[ch.~1]{smith1978evolution}. 
 
 \section{Genetics} 
\label{app:genetics}
 
In this appendix, we include more details about how our genetic system differs from yours. The important distinction to remember is that our species have a triploid genome that is made from combining three haploid gametes from our three parents (each of a distinct self-avoiding mating type), as opposed to your diploid genome that is made from combining two haploid gametes from your two parents.

\subsection*{Recombination}

In our species, each of our mating types makes a haploid gamete. The genetic material in this gamete is a third of the genetic material of its producer. Importantly, there is a recombination event that occurs before the production of the gametes to ensure that there is genetic variation in the gamete pool. Homologous chromosomes (i.e., chromosomes containing the same genes) are first aligned and bundled at their centers into a prism-like shape. Following this alignment event, several double strand breaks are made along each of the homologous chromosomes. Each broken strand is stabilized by the alignment of the other two chromosomes. The biological details of crossing-over between chromosomes are similar to that of those observed on Earth. Double strand breaks allow for invasion of another chromosome (see fig.~\ref{fig:crossnew} in sec.~\ref{sec:genetics}). The invading chromosome is determined by proximity to the broken chromosome in the prism, leading do the formation of a prism structure. This structure then proceeds by moving along all the strands and is then resolved by structure-selective enzymes. Each of the double strands breaks is resolved using the process detailed above. The chromosomal prism is then dissolved.       

\subsection*{Gamete formation and mating type \\ determination}
As discussed previously, our species is composed of three self-avoiding mating types. These mating types are distinguished by three sex-determining genotypes: `XXY,' `XYY,' and `XYZ.' To illustrate gamete formation in our species, we will first describe the simplest case, which is that of gamete formation in our `female' species (genotype `XYZ'). 

\begin{figure}
    \centering
    \includegraphics[width = 0.99\textwidth]{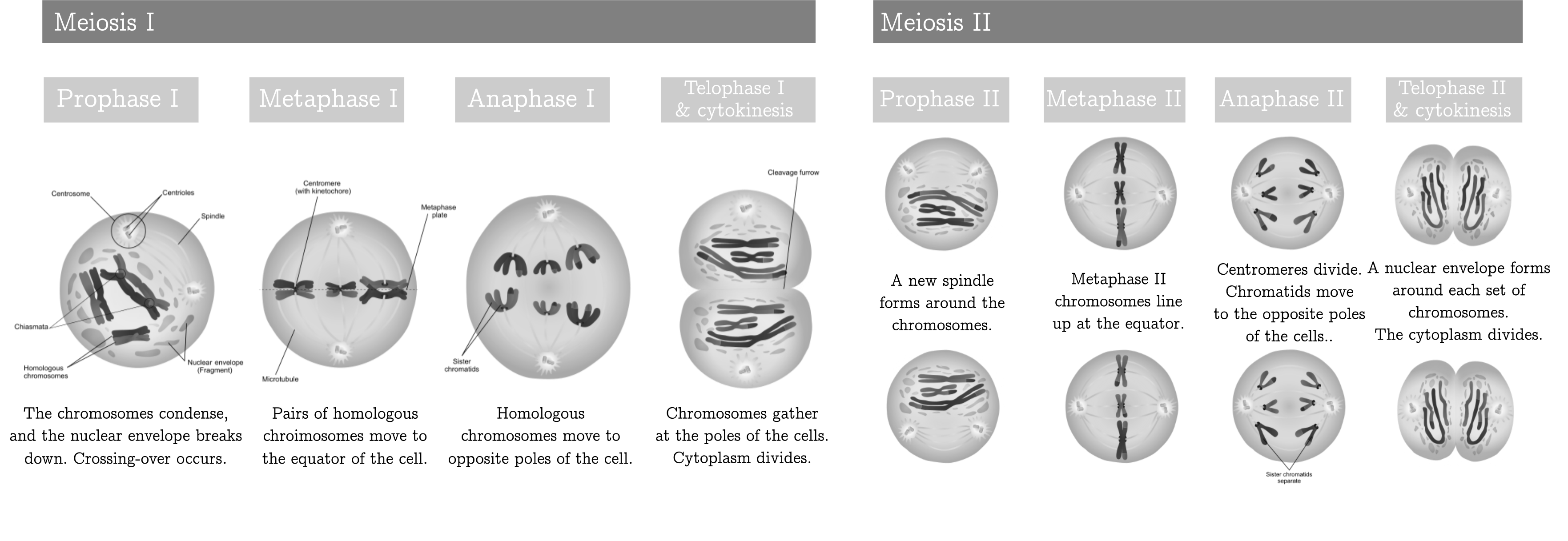}
    \caption{Diagram of the phases of meiosis in humans and similar Earth-based life. Adapted from: Ali Zifan, CCA-SA 4.0}
    \label{fig:meiosis_human}
\end{figure}

\begin{figure}
    \centering
    \includegraphics[width = 0.99\textwidth]{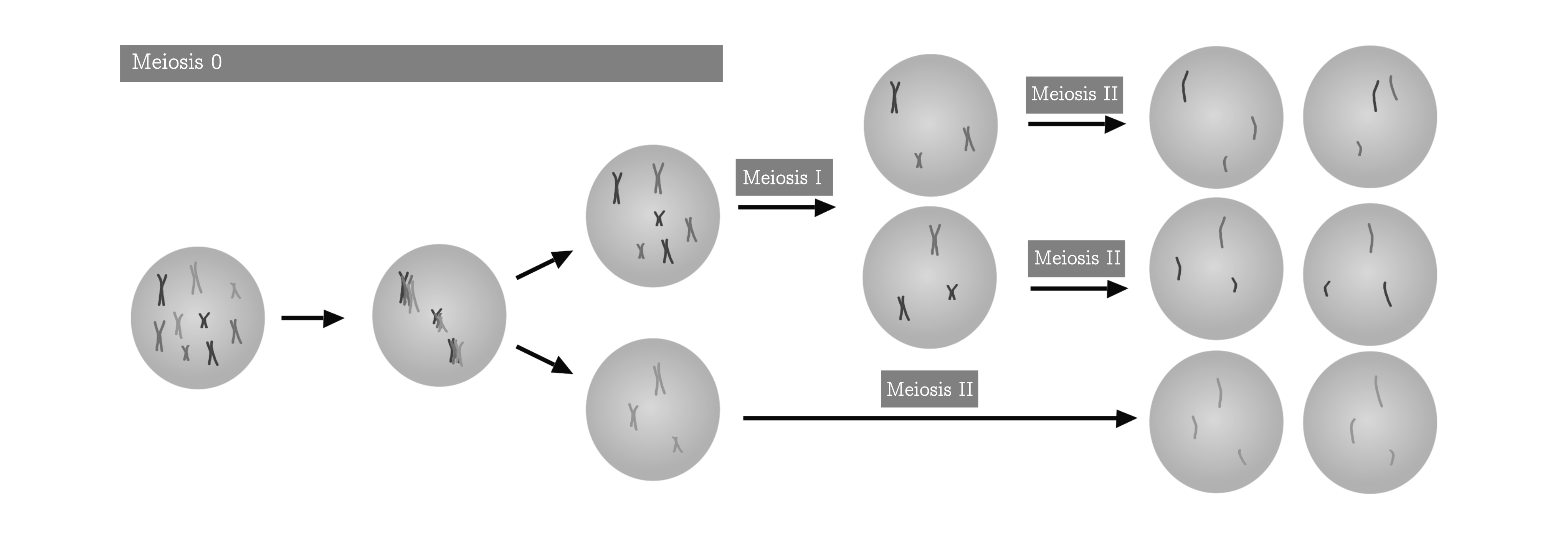}
    \caption{Diagram of the phases of meiosis in our species. Intra-chromosomal recombination was not depicted here for clarity.}
    \label{fig:meiosis_us}
\end{figure}

Once the recombination event occurs, a cellular event analogous to what you call meiosis occurs to form haploid gametes, see figure~\ref{fig:meiosis_human}. In the case of our species however, there is an asymmetrical meiosis phase that we will call `Meiosis 0,' see figure~\ref{fig:meiosis_us}, where the cells split in such a way that it forms a diploid daughter cell and a haploid daughter cell. The second step of our `Meiosis'-like event occurs only on the diploid cell and is analogous to your `Meiosis 1' step to form two haploid cells. All haploid cells then undergo a process analogous to `Meiosis 2' to form all the gametes.

\begin{figure}
    \centering
    \includegraphics[width = 0.9\textwidth]{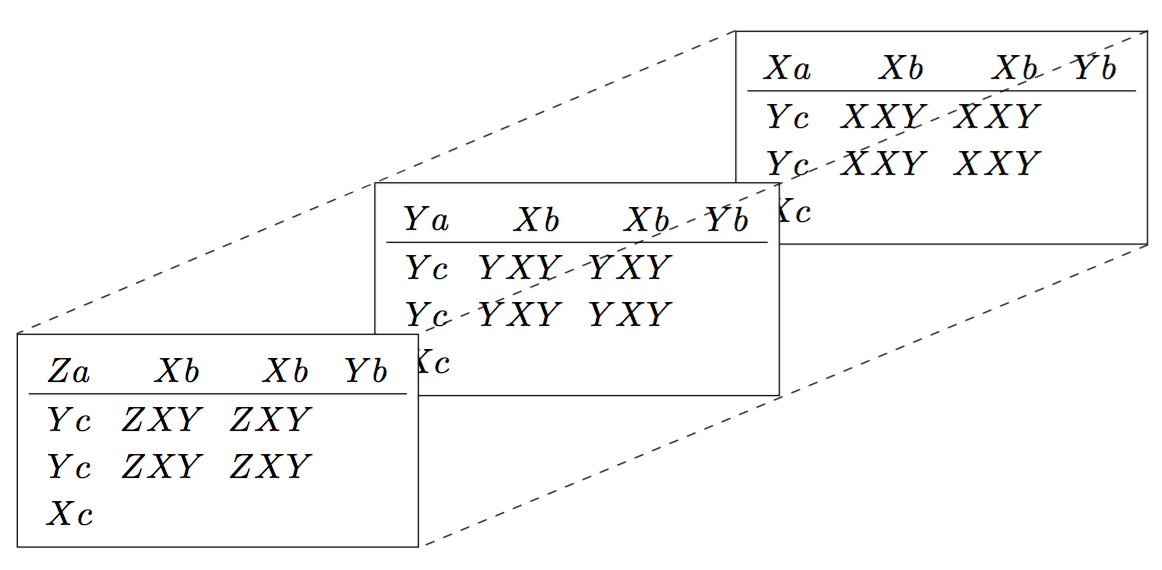}
    \caption{Partial Punnett cube of mating type assignment in our species. Letters $a$, $b$ and $c$ are used to help follow the three-mating types.}
    \label{fig:Punnett_Cube_SM}
\end{figure}

Crucially, this asymmetrical division, `Meiosis 0,' is \emph{non-random} in the two `male' mating types. The two `male' mating types (genotypes XXY and XYY) each have a pair of homologous sex-determining chromosomes (XX and YY respectively). In this case, unlike the fully heterologous XYZ female, the homologous pairs of chromosomes interact more strongly, causing them to always stay on the same side of the asymmetrical `Meiosis 0' step. This leads to the third sex-determining chromosome (Y and X respectively) to be separate and subsequently eliminated from the gamete formation process. In summary, genotype XXY only forms gametes with an X sex-determining chromosome while XYY only forms gametes with an Y sex-determining chromosome. The elimination of one of the sets of chromosomes at this stage is similar to that observed in a triploid toad species on your planet \autocite{stock2011simultaneous}). In the heterologous XYZ female (with respect to sex-determining chromosomes), there is no elimination of cells that lead to gametes and as such, there is an even distribution of gametes with X, Y and Z sex-determining chromosomes in the gametes. Fertilization then occurs when three gametes, each originating from a different mating type, meet and interact. The offspring's genetic material is thus a third of each of its parents. In conclusion, as illustrated in figure~\ref{fig:Punnett_Cube_SM}, the female's gamete determines the mating-type of the offspring, and all three mating-types are distributed evenly at birth.

\section{Gamete Formation and Mating Type Determination}
\label{app:gametes}

\begin{figure}[htb!]
    \centering
    \includegraphics[width = 0.9\textwidth]{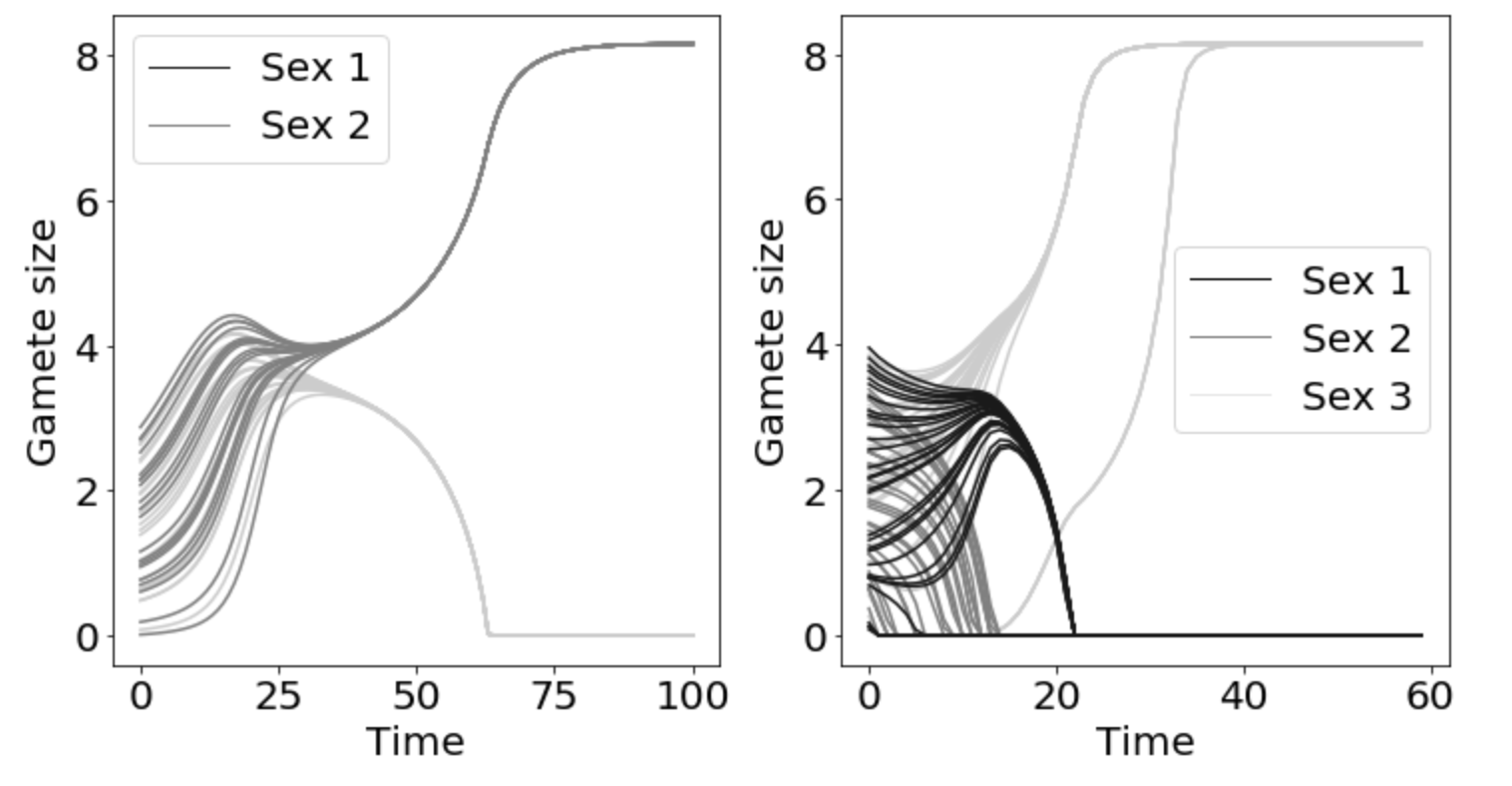}
    \caption{Simulation for the gamete size evolution. The shades of gray distinguish gametes of different sexes. Each curve indicates the evolution trajectory of one initial condition in gamete size. \textbf{Left:} in the case of two sexes, one sex evolves to have the minimal gamete size ($0$), and the other sex evolves to have a large gamete size. This recovers what happens on Earth. \textbf{Right:} in the case of three sexes, one sex evolves to have a large gamete size, and two evolves to have the minimal gamete size.}
    \label{fig:gameteDynamics}
\end{figure}

In our three-sex world, we have one sex with large gametes (analogous to eggs in humans) and two sexes with minimally small gametes (analogous to sperm in humans). To give you humans a more intuitive picture of what the three-sex world looks like, we present here a dynamical model and show, in simulation, demonstrate the gamete size evolution. 

\subsection*{Two sexes}
To convince you of the model's validity, we first present the model in the case of two sexes and show how the model predicts the emergence of gamete size differences on Earth. 

In the model, we use subscript 1 to denote variables for sex 1, and subscript 2 to denote those for sex 2. Let $x$ denote gamete size (biomass), and $\rho_1(x_1)$ denote the distribution of gamete size in sex 1, and $\rho_2(x_2)$ denote that in sex 2. Let there be $N$ individuals in either sex. We denote the gamete size of all individuals in sex 1 to be vector $X_1 = \langle x_1^{(1)}, x_1^{(2)}, \cdots,  x_1^{(N)} \rangle$. The superscripts denote the indexes for the individuals. Similarly, we denote that for sex 2 there is a vector $X_2 =\langle x_2^{(1)}, x_2^{(2)}, \cdots,  x_2^{(N)} \rangle$. 

We denote the minimum size of the gamete to be $x = 0$. We also denote the number of gametes an individual produces in one mating event to be $n$, which is expected to decrease with $x$. One function that satisfies this relationship is $n(x) = n_{max}/(x + 1)$, where $n_{max}$ is the maximum number of gametes possible produced in a mating event, when gametes are at its minimal size. This inverse relationship reflects the conservation of reproductive energy. The $1$ added to the denominator is for preventing singularity in this function. Consider a well-mixed population, where all individuals in sex 1 can mate with all individuals in sex 2, and vice versa. Consider one mating event, where an individual of sex 1, with gamete size $x_1$, mate with an individual of sex 2, with gamete size $x_2$ (we omit the superscript in the derivation for the simplicity). Note that an individual of sex 2 also mates with all other individuals in sex 1 at other times. Then the probability that one of $x_1$'s gamete merges with the $x_2$ gamete is, $n(x_1)/\sum_{x_i \in X_1}n(x_i)$. Given that the two gametes merge, the two merged gametes need to provide enough support in terms of biomaterial in order to produce children successfully. The probability of children's birth success increases with the biomass of the sum of the two merged gametes, with a function we call $S(x_1 + x_2)$. One choice of the function $S$ is the Sigmoid function, $S(x) = 1/(1 + \exp(-k (x - b)))$. 

In this mating event, the number of expected offsprings is the product of the two terms,  

\begin{equation}\label{eq:c1}
    c(x_1, x_2) = \frac{n(x_1)}{\sum_{x_i\in X_1} n(x_i)}S(x_1+ x_2)
\end{equation}
Integrate over all $x_2$'s, the expected number of total children for an individual of gamete size $x_1$ and sex 1 in all mating events is, 
\begin{equation}
    C_1(x_1) = \sum_{x_2 \in X_2} \frac{n(x_1)}{\sum_{x_i \in X_1} n(x_i)} S(x_1+ x_2) 
\end{equation}

Similarly, we can write down the number of expected total children for an individual of sex 2, with gamete size $x_2$, 
\begin{equation}
    C_2(x_2) = \sum_{x_1 \in X_1} \frac{n(x_2)}{\sum_{x_j\in X_2} n(x_j)} S(x_1+ x_2) 
\end{equation}

We assume, over evolutionary time,that individuals of both sexes evolve to increase the number of total expected children. We obtain a system of coupled ordinary differential equations (ODEs)

\begin{equation} \label{eq:2sexDynam}
\Bigg\{
\begin{array}{ll}
\frac{dx_1}{dt} = \frac{dC_1}{dx_1} \\
\frac{dx_2}{dt} = \frac{dC_2}{dx_2}  
     \end{array}
\end{equation}
The intuitive understanding of equation~\ref{eq:2sexDynam} is that gamete size evolves in the direction that increases the expected number of offsprings for individuals with that gamete size. Note that the collection $X_{1,2}$ changes over time. 

We numerically simulate the model (eq.~\ref{eq:2sexDynam}) with a finite, discrete population, with discrete time steps. We start with twenty individuals in each sex with gamete size drawn from a uniform distribution. The trajectories for the evolution of gamete sizes are shown in the left panel of figure~\ref{fig:gameteDynamics}. One sex evolves to have the smallest possible gametes size, and the other evolves to have a large gamete size, agreeing with what we observe on Earth.

\subsection*{Three sexes}

Now that we have confidence that the model predicts what happens on Earth, let us take a look at what the model shows in the case of three sexes (1, 2, and 3). 

In the world of three sexes, the number of children expected in an mating event with gamete sizes $x_1$, $x_2$, and $x_3$ (similar to eq.~\ref{eq:c1}) is modified to be
\begin{equation}\label{eq:c3_1}
    c(x_1, x_2, x_3) = \frac{n(x_1)}{\sum_{x_i\in X_1} n(x_i)} S(x_1+ x_2 + x_3)
\end{equation}
Intuitively, this equation combines the competition within the same sex, and cooperation with the other two sexes. Now we need to integrate $c$ over all mates in sex 2 and sex 3. The expected total number of children for an individual of sex 1 and gamete size $x_1$ is
\begin{equation} \label{eq:c3_2}
    C_1(x_1) = \sum_{x_2 \in X_2} \sum_{x_3 \in X_3}
   \frac{n(x_1)}{\sum_{x_i\in X_1} n(x_i)} S(x_1+ x_2 + x_3) 
\end{equation}
We believe you earth scientists are smart enough to write down expressions for $C_2$ and $C_3$, which are similar to equation~\ref{eq:c3_2}. Then the three-sex coupled dynamical system becomes a three-equation coupled ODE system 

\begin{equation} \label{eq:3sexDynam}
\Bigg\{
\begin{array}{ll}
\frac{dx_1}{dt} = \frac{dC_1}{dx_1} \\
\frac{dx_2}{dt} = \frac{dC_2}{dx_2}  \\
\frac{dx_3}{dt} = \frac{dC_3}{dx_3} \\
\end{array}
\end{equation}

Numerically simulating equation~\ref{eq:2sexDynam} using the same procedure as above (in discrete time step and discrete individuals) gives the evolutionary trajectories shown in the right panel of figure~\ref{fig:gameteDynamics}. One sex evolves to have a large gamete size, and the other two evolve to have the smallest possible gamete size.

\section{An Evolutionary Stable Strategy for Flower-Flier Species}
\label{app:ess}
Our dominant species has three distinct biological sexes, unlike yours, \emph{Homo sapiens}, which has two. As with \emph{H. sapiens}, there exists asymmetry and this asymmetry has important consequences for the physical traits of our (and your) species, which in turn has consequences for the organization of our society. Here, we provide a model of gamete asymmetry as an Evolutionarily Stable Strategy (ESS).

Let $m_F$, $m_P$, and $m_C$ be the gamete sizes of our Flower, Pollinator, and Central species. We assume a minimum viable gamete size, $m_0$, and represent the ESS gamete sizes by  $m^*_F$, $m^*_P$, and $m^*_C$. The overall viability, $b$, of the double fertilized ``egg'' is a function of the sum of the gamete sizes, $b=b(m_F,m_P,m_C)$. There are three relevant functions, corresponding to a mutant of each sex attempting to invade the ESS, $W_F(m_F,m_F^*,m_P^*,m_C^*)$, $W_P(m_P,m_F^*,m_P^*,m_C^*)$, and $W_C(m_C,m_F^*,m_P^*,m_C^*)$. While this could be analyzed using the Karush–Kuhn–Tucker (KKT) conditions, the problem is simple enough to intuit the relevant inequalities to enforce where constraints---notably boundary conditions---exist.

For simplicity, we assume that each $P$ and $C$ individual follows either a single- or double-mating strategy, and that the probability of each sex adopting a double-mating strategy is, respectively, $q^{(P)}$ and $q^{(C)}$. Given this, each ``mating'' of an $F$ individual faces fertilization competition with probability $p^{(P)} = \frac{2 q^{(P)}}{1+q^{(P)}}$ and each mating of a $P$ individual faces fertilization competition with probability $p^{(C)} = \frac{2 q^{(C)}}{1+q^{(C)}}$. We first consider the ESS condition for $F$ individuals, who adopt a gamete size near the minimum viable size.

Consider now the branching probabilities in table~\ref{tab:branching} for ``matings.'' Whereas increasing gamete size improves the viability of offspring, it reduces the amount of sperm competing in each mating, and we assume that the weighting in this competition is inversely proportional to the gamete size \autocite[c.f.][]{parker1998}. Hence, the fitness of an $F$ mutant following strategy $m_F$ while the others follow $m_P^*$ and $m_C^*$ is

\begin{flalign}
  \label{eq:W_F}
  W_F(m_F,m_F^*,m_P^*,m_C^*) &= b(m_F,m_P^*,m_C^*) \, \, * & \nonumber \\ 
  \big[& \nonumber &\\
  (1-p^{(C)}) \, (1-p^{(P)})& \, \, \nonumber & \\
  +(1-p^{(C)}) \, p^{(P)}& \, \, \omega(m_F,m_F^*,1) \nonumber & \\
  +p^{(C)} \, (1-p^{(P)}) \, (1-p^{(P)})& \, \, \omega(m_F,m_F^*,1) \nonumber & \\
  +p^{(C)} \, (1-p^{(P)}) \, p^{(P)}& \, \, \omega(m_F,m_F^*,2) \nonumber & \\
  +p^{(C)} \, p^{(P)} \, (1-p^{(P)})& \, \, \omega(m_F,m_F^*,2) \nonumber & \\
  +p^{(C)} \, p^{(P)} \, p^{(P)} & \, \, \omega(m_F,m_F^*,3) \nonumber & \\
  &\big]&
\end{flalign}

\noindent where for convenience we define the function $\omega$,
\begin{align*}
\omega(m,m^*,\alpha) = \frac{ \frac{1}{m} }{ \frac{1}{m}+\frac{\alpha}{m^*} }.
\end{align*}

\noindent The first derivative of $\omega$ with respect to $m$ is
\begin{align*}
    \frac{\partial \omega}{\partial m} &= \frac{ -\frac{1}{m^2} }{ \frac{1}{m}+\frac{\alpha}{m^*} }+ \frac{ \frac{1}{m^3} }{ [\frac{1}{m}+\frac{\alpha}{m^*}]^2 }\\
    &=-\frac{1}{m}\omega(m,m^*,\alpha)(1-\omega(m,m^*,\alpha)).
\end{align*}

\noindent Evaluating $\omega$ at $m=m^*$ yields

\begin{align*}
    \omega(m^*,m^*,\alpha) = \frac{1}{1+\alpha}.
\end{align*}
Therefore,
\begin{align*}
  \frac{\partial \omega}{\partial m}|_{m=m^*} &=  -\frac{\alpha}{m^*(1+\alpha)}. 
\end{align*}

The first derivative of the $W_F$ with respect to $m_F$ evaluated at $m_F=m_F^*$ is
\begin{align}
    \label{eq:cond_F}
    \frac{\partial W}{\partial m_F}|_{m_F=m_F^*} =
    k_1 \, b'(m^*) - \frac{1}{m_F^*} \, k_2 \, b(m^*)
\end{align}

\noindent where

\begin{equation}
  k_1 = 1 - \frac{p_P + p_C}{2} + \frac{1}{6} p_P p_C + \frac{1}{12} p_P^2 p_C
\end{equation}

\noindent and

\begin{equation}
  k_2 = \frac{p_P + p_C}{2} - \frac{1}{6} p_P p_C - \frac{1}{12} p_P^2 p_C \mbox{.}
\end{equation}

\begin{center}
\begin{tabular}{ |r|r|r|c|c|c| } 
 \hline
             &             && $N_C$ & $N_P$ & $N_F$ \\ 
 $1-p^{(C)}$ & $1-p^{(P)}$ && 1     & 1     & 1 \\ 
 $1-p^{(C)}$ &   $p^{(P)}$ && 1     & 1     & 2 \\ 
   $p^{(C)}$ & $1-p^{(P)}$ & $1-p^{(P)}$ & 1     & 2     & 2 \\ 
   $p^{(C)}$ & $1-p^{(P)}$ & $p^{(P)}$ & 1     & 2     & 3 \\ 
   $p^{(C)}$ & $p^{(P)}$ & $1-p^{(P)}$ & 1     & 2     & 3 \\ 
   $p^{(C)}$ & $p^{(P)}$ & $p^{(P)}$ & 1     & 2     & 4 \\ 
 \hline
\end{tabular}
\label{tab:branching}
\end{center}

\noindent Since $m_F^*=m_0$ at the ESS, the following inequality constraint must be satisfied:
\begin{equation}
  \label{eq:W_F_ineq}
    \frac{\partial W_F}{\partial m_F}|_{m_F=m_F^*} < 0 \mbox{.}
\end{equation}
Invoking the so-called Marginal Value Theorem \autocite{charnov1976,parker1998}, the optimal investment in offspring viability is achieved when the following condition on overall gamete size is satisfied:

\begin{equation}
  \label{eq:mvt}
  b'(m_F^*+m_P^*+m_C^*) = \frac{b(m_F^*+m_P^*+m_C^*)}{m_F^*+m_P^*+m_C^*} \mbox{.}
\end{equation}

\noindent Combining equations~\ref{eq:cond_F}, \ref{eq:W_F}, and~\ref{eq:mvt}, the optimal gamete size for $F$ is $m_F^*=m_0$ so long as the following inequality is satisfied:

\begin{equation}
  \label{eq:inequa_F}
  \frac{m_F^*}{m_F^*+m_P^*+m_C^*} < \frac{k_2}{k_1} = \frac{\frac{p_P + p_C}{2} - \frac{1}{6} p_P p_C - \frac{1}{12} p_P^2 p_C }{ 1 - \frac{p_P + p_C}{2} + \frac{1}{6} p_P p_C + \frac{1}{12} p_P^2 p_C} \mbox{.}
\end{equation}

\noindent A similar derivation can be applied to $P$, but with the number $N_P$ standing in for $N_F$ in Table~\ref{tab:branching}. One conclude that

\begin{equation}
  \label{eq:inequa_P}
  \frac{m_P^*}{m^*} < \frac{k_2}{k_1} = \frac{1}{\frac{2}{p_C}-1} \mbox{.}
\end{equation}

\section{Matching/Unmatching}
\label{app:matchingdivorce}

Here we present a simple model based on a concept you seem to be comfortable with, the ``Nash equilibrium.'' The population consists of $N$ types of sex, and each sex contains $n$ number of agents. Let $a_{i,s}$ be an agent ID $i$ from sex $s$. Each agent is characterized by its feature $X_{i,s}$. They can form a union by find a mate from each other sex.  Their preference for union is represented by a utility function $u(\{a_{j,t} \}_{t\neq s}  ;a_{i,s})$ representing the utility of an agent $a_{i,s}$ union with a set of agents $\{a_{j,t} \}_{t\neq s}$.

For each period, we repeat the following steps:
\begin{enumerate}
    \item Pick up one person from each sex randomly.
    \item They three consider the utility from their union.
    \item If {\textit all} of them think this new union is better than their current one, then they each break up the existing union and form the new one.
    \item Else, nothing happens.
\end{enumerate}
As this is a best-response dynamics, the following proposition trivially holds:
\begin{proposition}
There exists a fixed point in this union process.
\end{proposition}

\section{Cultural Mixing under Triparental Reproduction}

\label{app:culturalmixing}

We provide a probabilistic model for calculating an individual's expected number of cultural affinities, such as language or membership in a legal forming group, and show how it provides evidence for an explanatory role for three-parent union in the emergence of cultural homogeneity in our society. Consider a population of individuals $I$. Let $X_{i}$ be an integer-valued random variable that denotes the number of cultural affinities with which an individual $i\in I$ is born. Let $Y_{i}$ be an integer-valued random variable whose value is denotes the number of distinct cultures with which $i$'s union partners have cultural affinities. $Y_{i}=0$ if and only if the individual $i$ does not form any unions or only forms unions with those whose cultural affinities are a subset of the cultural affinities with which $i$ is born. Finally, let $Z_{i}$ be an integer-valued random variable whose value is defined via the following equation:
\begin{equation}\label{sum}
    Z_{i}=X_{i} + Y_{i}
\end{equation}
In other words, the value of $Z_{i}$ denotes $i$'s total set of distinct cultural affinities, acquired either by birth or through union. Each of these four random variables is measurable with respect to a probability space $(\Omega,\mathcal{P}(\Omega),P(\cdot))$, where $\Omega$ is the set of all possible $n$-individual populations and $\mathcal{P}(\Omega)$ is the power set of $\Omega$, i.e.\ the set of all subsets of $\Omega$, plus the empty set.\par

Assume that $X_{i}$ for each individual $i$ is independently drawn from an identical probability distribution. The expected number of cultural affinities for each individual, assuming that there are $n$ possible cultural affinities is calculated as follows: 
\begin{equation}
    \mathbbm{E}(Z_{i})=\sum_{\alpha=0}^{n}\alpha P(Z_{i}=\alpha)
\end{equation}
From equation (\ref{sum}), we have:
\begin{equation}
    \mathbbm{E}(Z_{i}) =  \sum_{\alpha=0}^{n}\alpha P(X_{i}+Y_{i}=\alpha)
\end{equation}
Letting $Y_{i}=\beta$ we expand the expression as follows:
\begin{equation}
    \mathbbm{E}(Z_{i}) =  \sum_{\beta=0}^{n}\sum_{\gamma=0}^{n}(\beta+\gamma) P(Y_{i}=\beta|X_{i}=\gamma)P(X_{i}=\gamma)
\end{equation}
Based on historical data from our star system, during periods of cultural heterogeneity the probability $P(X_{i}=\gamma)$ is best estimated via the binomial probability mass function:
\begin{multline}
    P(X_{i}=\gamma) = f(\gamma,\mu_{X_{i}}) \\ = \ {n\choose\gamma}\bigg(\frac{\mu_{X_{i}}}{n}\bigg)^{\gamma}\bigg(1-\frac{\mu_{X_{i}}}{n}\bigg)^{n-\gamma}
\end{multline}
Here, $\mu_{X_{i}}$ denotes the mean value of $X_{i}$. The value of the conditional probability $P(Y_{i}=\beta|X_{i}=\gamma)$, for any $\alpha$ and $\gamma$, is best estimated by the following following binomial probability mass function:
\begin{multline}
    P(Y_{i}=\beta|X_{i}=\gamma) =  g(\beta,\gamma) \\ = \ {n-\gamma\choose\beta}\bigg(\frac{.05+\gamma}{n}\bigg)^{\beta}\bigg(1-\frac{.05+\gamma}{n}\bigg)^{n-\gamma-\beta}
\end{multline}
Note that the mean number of distinct cultural affinities of a person's union partners increases only slightly with the cultural affinities with which one is born. This reflects the fact that in our society, as on Earth, union partners tend to share cultural affinities \autocite[see][]{Schwartz2013}.\par

These findings allow us to re-write the equation for $\mathbbm{E}(Z_{i})$ as follows:\ 
\begin{equation}
    \mathbbm{E}(Z_{i})=\sum_{\gamma=0}^{n}\sum_{\beta=0}^{n}(\beta+\gamma) g(\beta,\gamma)f(\gamma,\mu_{X_{i}})
\end{equation}
Note that this renders the expectation $\mathbbm{E}(Z_{i})$ a function solely of the mean number of cultural affinities that a person is born with, i.e.\ $\mu_{X_{i}}$. As shown in figure~\ref{fig:cultureplot}, $\mathbbm{E}(Z_{i})$ is consistently greater than $\mu_{X_{i}}$ for various values of $\mu_{X_{i}}$ and $n$, and increases nearly linearly with $\mu_{X_{i}}$, with slope greater than 1, for low values of $\mu_{X_{i}}$ relative to $n$. Given that individuals inherit most of their parents' cultural affinities, this leads to an increase in the mean number of cultural affinities held by each member of the population over time, until (we hypothesize) it reaches an unmanageable level, such that a monoculture becomes a preferable alternative. {\glyphabet{E}}





\end{mainmatter}


\begin{backmatter}

\chapter[Bibliography][]{}
\chaptermark{Bibliography} 
\addcontentsline{toc}{section}{} 
\vskip\afterchapskip 

\pagestyle{bibheadings} 

\printbibliography[title={Bibliography}]



\end{backmatter}


\end{document}